%% file: fiberpath_scf.tex
\documentclass[sigconf]{acmart}

\usepackage{booktabs} 

\usepackage[ruled]{algorithm2e} 

\SetAlFnt{\small}
\SetAlCapFnt{\small}
\SetAlCapNameFnt{\small}
\SetAlCapHSkip{0pt}

\input{packages}
\input{macros}

\copyrightyear{2023} 
\acmYear{2023} 
\setcopyright{rightsretained} 
\acmConference[SCF '23]{Symposium on Computational Fabrication}{October 8--10, 2023}{New York City, NY, USA}
\acmBooktitle{Symposium on Computational Fabrication (SCF '23), October 8--10, 2023, New York City, NY, USA}
\acmDOI{10.1145/3623263.3623356}
\acmISBN{979-8-4007-0319-5/23/10}

\begin{document}
\title{More Stiffness with Less Fiber: End-to-End Fiber Path Optimization for 3D-Printed Composites}

\author{Xingyuan Sun}
\affiliation{
 \institution{Princeton University}
 \city{Princeton}
 \state{NJ}
 \country{USA}
}
\email{xs5@princeton.edu}
\author{Geoffrey Roeder}
\affiliation{
 \institution{Princeton University}
 \city{Princeton}
 \state{NJ}
 \country{USA}
}
\email{roeder@princeton.edu}
\author{Tianju Xue}
\affiliation{
 \institution{Hong Kong University of Science and Technology}
 \city{Hong Kong}
 \country{China}
}
\email{cetxue@ust.hk}
\author{Ryan P. Adams}
\affiliation{
 \institution{Princeton University}
 \city{Princeton}
 \state{NJ}
 \country{USA}
}
\email{rpa@princeton.edu}
\author{Szymon Rusinkiewicz}
\affiliation{
 \institution{Princeton University}
 \city{Princeton}
 \state{NJ}
 \country{USA}
}
\email{smr@princeton.edu}

\input{text/abstract}



\maketitle

\input{text/intro}
\input{text/related_work}
\input{text/method}
\input{text/hardware}
\input{text/modulus}
\input{text/experiments}
\input{text/ablation}
\input{text/discussion}
\input{text/ack}

\bibliographystyle{ACM-Reference-Format}
\bibliography{fiberpath_scf}

\clearpage
\input{text/appendix}

\end{document}

%% file: packages.tex
\usepackage{color,xcolor}

\usepackage{colortbl}
\usepackage{float,wrapfig}
\usepackage{multirow}
\usepackage{subcaption}

\usepackage{amsfonts,amsthm}
\usepackage{bm}
\usepackage{bbold}
\usepackage{microtype}

\usepackage{soul}
\usepackage{xspace}

\usepackage{url}

\usepackage{algorithmic}
\usepackage{mathtools}

%% file: macros.tex
\newcommand{\sect}[1]{Section~\ref{#1}}
\newcommand{\ssect}[1]{\S~\ref{#1}}
\newcommand{\eqn}[1]{Equation~\ref{#1}}

\newcommand{\fig}[1]{Figure~\ref{#1}}

\newcommand{\tbl}[1]{Table~\ref{#1}}

\newcommand{\ignore}[1]{}

\makeatletter
\DeclareRobustCommand\onedot{\futurelet\@let@token\@onedot}
\def\@onedot{\ifx\@let@token.\else.\null\fi\xspace}

\def\eg{\emph{e.g}\onedot} 
\def\ie{\emph{i.e}\onedot} 
 
\def\etc{\emph{etc}\onedot}

\makeatother

%% file: text/abstract.tex
\begin{abstract}

In 3D printing, stiff fibers (\eg, carbon fiber) can reinforce thermoplastic polymers with limited stiffness.
However, existing commercial digital manufacturing software only provides a few simple fiber layout algorithms, which solely use the geometry of the shape.
In this work, we build an automated fiber path planning algorithm that maximizes the stiffness of a 3D print given specified external loads. We formalize this as an optimization problem: an objective function is designed to measure the stiffness of the object while regularizing certain properties of fiber paths (\eg, smoothness).
To initialize each fiber path, we use finite element analysis to calculate the stress field on the object and greedily ``walk'' in the direction of the stress field.
We then apply a gradient-based optimization algorithm that uses the adjoint method to calculate the gradient of stiffness with respect to fiber layout.
We compare our approach, in both simulation and real-world experiments, to three baselines: (1) concentric fiber rings generated by Eiger, a leading digital manufacturing software package developed by Markforged, (2) greedy extraction on the simulated stress field (\ie, our method without optimization), and (3) the greedy algorithm on a fiber orientation field calculated by smoothing the simulated stress fields.
The results show that objects with fiber paths generated by our algorithm achieve greater stiffness while using less fiber than the baselines---our algorithm improves the Pareto frontier of object stiffness as a function of fiber usage.
Ablation studies show that the smoothing regularizer is needed for feasible fiber paths and stability of optimization, and multi-resolution optimization helps reduce the running time compared to single-resolution optimization.

\end{abstract}

%% file: text/intro.tex
\section{Introduction}

\input{figText/teaser}

Additive manufacturing has revolutionized the ability to fabricate three-dimensional objects of high geometric complexity, with a variety of applications including in healthcare, automotive, and aerospace industries~\citep{shahrubudin2019overview}.
However, the increasing flexibility in manufacturing has outstripped our ability to produce designs that optimally take advantage of 3D printers.
This has motivated research on \emph{computational fabrication} pipelines that augment human specification of goals with computational optimization of designs that best realize those goals, for problems ranging from ensuring structural integrity through controlling appearance and fine-tuning the fabrication process~\cite{Attene18}.

In this work, we address the problem of producing structurally-sound parts that are capable of bearing nontrivial load.
We aim to exploit the capabilities of devices such as the Markforged Mark Two~\cite{MarkTwo}, which is based on conventional fused deposition modeling (FDM) using thermoplastic nylon, but augments this with the ability to extrude and deposit continuous fibers.
Options for the latter include carbon fiber, Kevlar, fiberglass, and HSHT (High Strength High Temperature) fiberglass, all of which offer the ability to selectively strengthen printed parts with respect to tensile loads.
In effect, this process creates fiber-reinforced plastic (FRP) composites~\citep{kabir2020critical}, but with the ability to control fiber placement to achieve specific tradeoffs in strength, weight, and cost.

The optimization of fiber layout is similar to problems traditionally considered in computational fabrication, such as topology optimization (\ie, removing material from certain regions) and spatially-varying assignment of different materials.
Systems for these latter tasks are typically based on Eulerian analysis and optimization, in which a quantity (density, material choice, \etc) is determined for each location in space (\eg, on a voxel grid). 
Similarly, almost all existing methods for optimizing carbon fiber composites focus on the spatially-varying fiber direction field, then use variants of greedy extraction or ODE solvers to extract the fiber paths themselves~\citep{wang2021load, schmidt2020structural}.

In contrast, we are inspired by a Lagrangian point of view:  we characterize the strength of the part as a function of the fiber path, compute gradients with respect to changes in path coordinates, and optimize the fiber path directly using gradient descent.
This strategy is based on the adjoint method~\citep{errico1997adjoint, cao2003adjoint}, commonly used for PDE-constrained optimization, and exploits modern systems for automatic differentiation~\citep{griewank2008evaluating}, which have evolved considerably in recent years to support a range of machine learning and general optimization problems. 
Our end-to-end optimization approach has the benefit of focusing directly on the final goal---maximizing stiffness with respect to external loads---rather than on indirect objectives such as minimizing strain throughout the object.

We incorporate our gradient descent-based optimization into a complete system that addresses three key challenges:
(1) solving for the stress field of the object given external loads,
(2) computing an optimization objective and its gradient based on the stress field, and
(3) providing a good initialization of fiber layout for our local optimizer.
To address the first challenge, we model the composite material using the linear elastic model, and approximately solve the PDE using the finite element method.
Without loss of generality and for the sake of simplicity, we model the composite material in two dimensions under the assumption of in-plane stress (\ie, we only consider laminates).
We also simplify the problem by considering Dirichlet (fixed-displacement) boundary conditions.
To address the second challenge, we design an objective function based on total strain energy given the boundary conditions: under the assumption of linear elasticity, maximizing this energy is equivalent to maximizing the object's stiffness.
We regularize the objective to ensure that the optimized fiber paths are feasible.
Finally, to address the last challenge, we initialize each fiber path by greedily following the directions of maximum tensile stress (or perpendicular to the direction of maximum compressive stress). We further use a multi-resolution approach inspired by multigrid methods, to reduce the running time of the optimization.

We show designs produced by our method on a number of illustrative case studies, demonstrating that our method yields higher stiffness with less fiber as compared to baseline paths produced by the Eiger software by Markforged~\shortcite{Eiger}.
We compare our results to greedy extraction based on either the stress field or optimized fiber direction field, as well as other ablations including omitting regularization or multi-resolution optimization.
We print our designs (see Figure~\ref{fig:teaser}), using the method of~\citet{Sun:2021:ASO} to compute fiber extruder paths that compensate for fiber stiffness.
Finally, we test our printed parts to verify that our method matches the predicted stiffness in the real world (subject to inherent print-to-print variations in material strength).

%% file: figText/teaser.tex
\begin{figure*}[t]
    \captionsetup[subfloat]{margin=7pt,format=hang,singlelinecheck=false}
    \centering
    \begin{subfigure}{0.24\linewidth}
        \includegraphics[width=\linewidth]{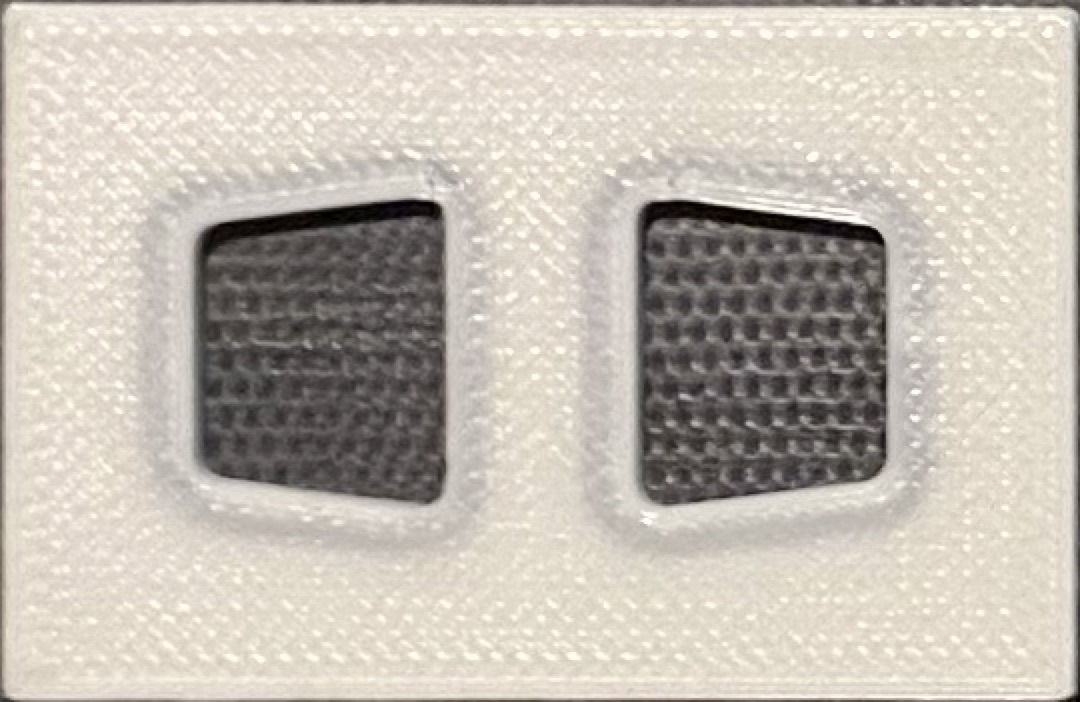}
        \caption{Eiger baseline (1 inner ring), \\ length = 449.6 mm, \\ stiffness = 292.3 N/mm.}
    \end{subfigure}
    \begin{subfigure}{0.24\linewidth}
        \includegraphics[width=\linewidth]{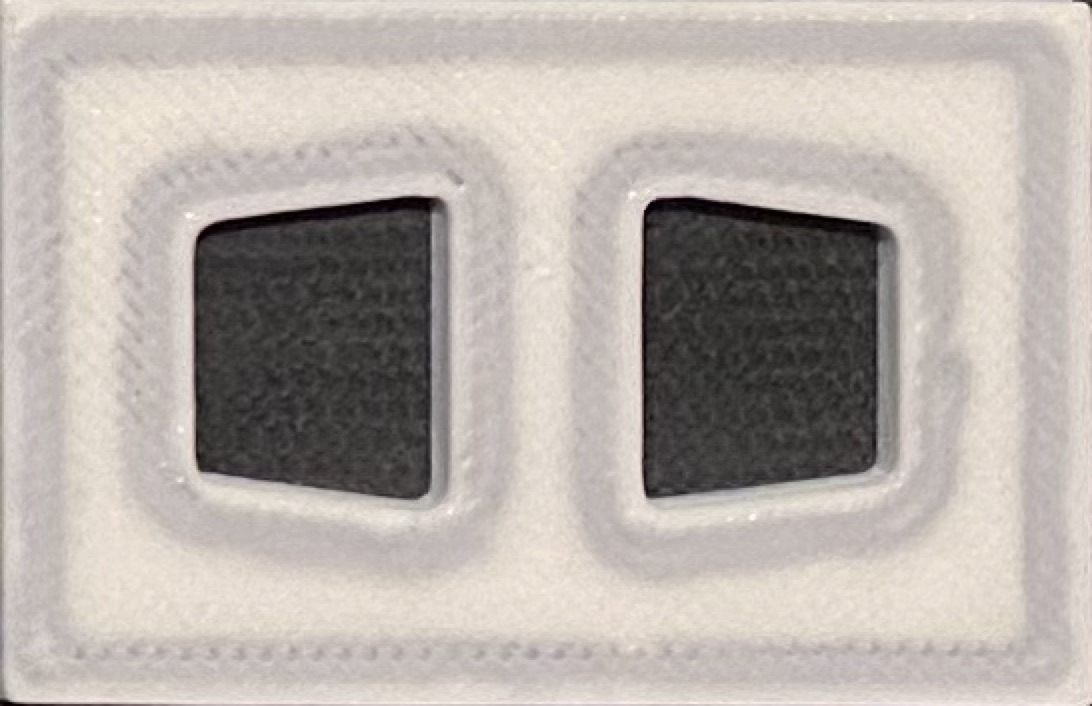}
        \caption{Eiger baseline (2 rings at all walls), \\ length = 2022.8 mm, \\ stiffness = 617.2 N/mm.}
    \end{subfigure}
    \begin{subfigure}{0.24\linewidth}
        \includegraphics[width=\linewidth]{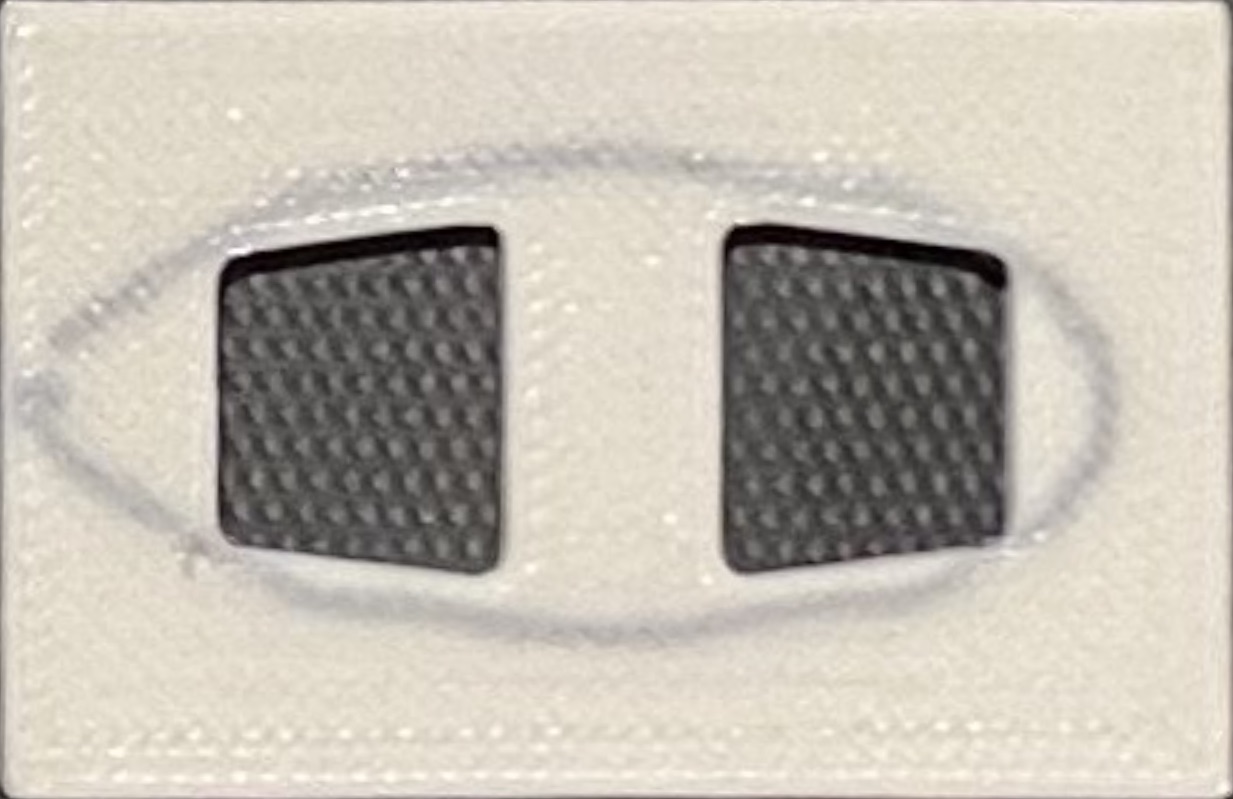}
        \caption{Optimized fiber path (1 ring), \\ length = 372.7 mm, \\ stiffness = 483.8 N/mm.}
    \end{subfigure}
    \begin{subfigure}{0.24\linewidth}
        \includegraphics[width=\linewidth]{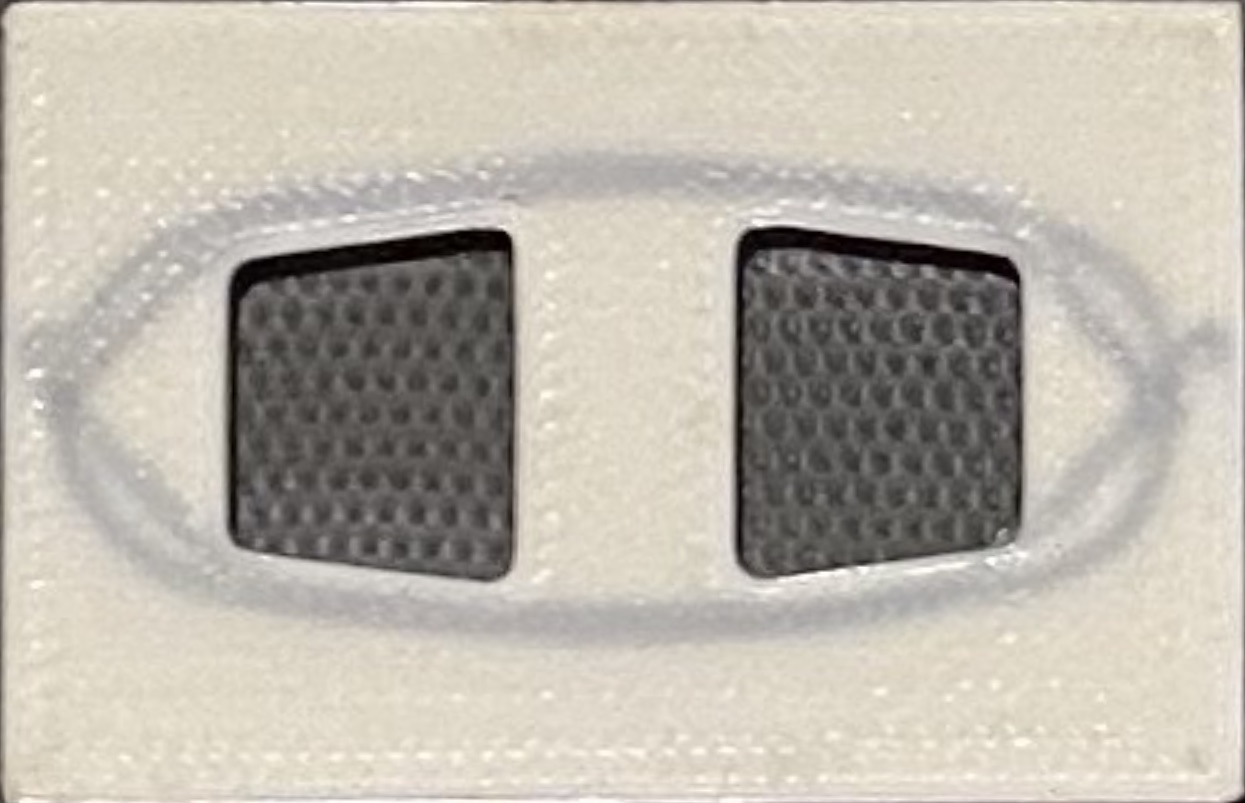}
        \caption{Optimized fiber path (2 rings), \\ length = 799.5 mm, \\ stiffness = 815.0 N/mm.}
    \end{subfigure}
    \caption{Planned and 3D printed fiber paths with fiber lengths and average stiffness measured over four batches annotated, for a part with external tension applied between two holes. (a) (b): Concentric fiber rings generated by the Eiger baseline only consider geometry. (c) (d): Our optimized fiber paths, tuned for the applied loads, yield greater stiffness at lower fiber lengths.}
    \label{fig:teaser}
\end{figure*}

%% file: text/related_work.tex
\section{Related work}

\subsection{Fiber orientation optimization in 3D printing}
\label{sec:prev_orientation}
A task that is similar to fiber path planning is fiber orientation optimization, where researchers discretize space into elements and optimize fiber orientations in them.
Additional steps, such as greedy extraction, ODE solvers, or geometric methods, must be performed to produce fiber paths from the orientation field.
Thus fiber orientation optimization can serve as the first step of fiber path planning, which we will discuss in~\ssect{sec:rlt-path-planning}.
See \citet{hu2021review} for a survey (called ``free material optimization'').
The most common approach for orientation optimization is to set density and orientation as design variables and optimize an objective such as compliance~\citep{chu2021robust, da2020topology, jung2022inverse} or the Tsai–Wu failure criterion~\citep{ma2022strength} with a gradient-based optimizer.
To address the checkerboard pattern issue (periodicity of the orientation variables), researchers usually use filtering~\citep{andreassen2011efficient} to smooth the design variable field (\eg, through a weighted average of neighboring elements).
Another choice of design variable is the lamination parameters: \citet{shafighfard2019design} and \citet{demir2019design} proposed to first optimize lamination parameters, search for the best fitting fiber orientations from the optimized lamination parameters, and then perform an optimization on the orientation field while considering manufacturing constraints (\eg, curvature).
There are also iterative variants.
For example, \citet{caivano2020topology} proposed iterating between calculating the principal stress direction and updating the material distribution until convergence.
While mainly concentrating on orientation optimization, some approaches do ultimately generate fiber paths.
For example, \citet{fedulov2021optimization} first optimized density and orientation and then used third-party software for printing trajectory generation;
\citet{schmidt2020structural} performed density and orientation optimization and generated streamlines using the 4th-order Runge-Kutta integrator for visualization.

\subsection{Fiber path planning in 3D printing}
\label{sec:rlt-path-planning}

A variety of path planning algorithms have been proposed for continuous fiber-reinforced plastics---see \citet{zhang2020review} for a survey.
The most common approach is to first perform an optimization (topology, orientation, \etc), and then extract fiber paths from the result.
As discussed, orientation optimization is one choice of the optimization (\ie, use density and orientation as the design variables), but there are different methods for path extraction.
\citet{wang2021load} proposed to ``walk'' in the field along with the stress direction and consider the angle turned in every move to produce smoothed paths.
\citet{papapetrou2020stiffness} described three methods for path extraction: the offset method and the EQS (Equally-Space) method use the geometry of the optimized layout, and the streamline method fits the orientation field with streamlines.
\citet{safonov20193d} proposed to alternate between topology optimization and fiber orientation updates using an evolutionary heuristic method.

There are also more potential choices for the design variable.
For example, one choice is to only optimize the density.
\citet{li2021full} performed topology optimization of material density (without orientation) using regularizers that force the fiber material to form lines.
However, they did not extract fiber paths explicitly at the end, so it is unclear whether the fibers are directly printable. 
\citet{li2020path} and \citet{chen2021topological} proposed to lay fibers along with the load transmission trajectories.
\citet{almeida2019cross} proposed to perform the SIMP (Solid Isotropic Material with Penalization) method first, designed the fiber pattern manually, and then used a genetic algorithm to determine the number of fiber rings/paths that would minimize compliance (defined as mass divided by stiffness).
\citet{sugiyama20203d} proposed to calculate the stress field and update fiber paths so that they follow the direction of maximum principal stress, repeating this process until convergence. 
Apart from these two-stage approaches, there are also end-to-end approaches based on genetic algorithms.
For example, \citet{yamanaka2016fiber} modeled fiber paths as streamlines and optimized them directly using a genetic algorithm.

In summary, most existing works perform fiber planning in two stages (topology/orientation optimization followed by path extraction).
In contrast, our method performs an end-to-end optimization of the fiber layout, maximizing the regularized object stiffness via a gradient-based optimizer.

\subsection{PDE-constrained optimization}
Also related to the problem of optimizing geometry to maximize stiffness is the area of PDE-constrained optimization, in which an optimization problem is subjected to physical constraints expressed via partial differential equations (PDEs)~\citep{de2015numerical}.
There are two common types of algorithms to solve PDE-constrained optimization problems: \textit{all-at-once} and \textit{black-box}~\citep{herzog2010algorithms}.
\textit{All-at-once} treats both the design variable and the state variable as independent variables, so the method may not satisfy the constraints before the optimization finishes.
A common \textit{all-at-once} algorithm is SQP (sequential quadratic programming)~\citep{boggs1995sequential}. 
A disadvantage of the \textit{all-at-once} approach is the dimension of the state variable can be very large, which makes the optimization costly.
\textit{Black-box} solves the problem in reduced form, by treating the design variable as the only independent variable, so that a gradient-based optimizer can be applied (\eg, gradient descent, Newton's method).
We formalize the fiber path planning task as a PDE-constrained optimization problem and use the \textit{black-box} approach, specifically the adjoint method, to solve it.

%% file: text/method.tex
\section{Method}
\label{sec:method}
\input{figText/pipeline}

The pipeline of our method is shown in~\fig{fig:pipeline}.
Starting from a goal (a shape with some external loads), we first simulate the stress field using the finite element method (\ssect{sec:method-simulation}).
We apply a greedy fiber extraction algorithm by ``walking'' in the stress field, and then downsample the greedy path (\ssect{sec:method-greedy}).
We build and optimize an objective function based on the object's stiffness and regularity conditions of the fiber paths, using the adjoint method to calculate the gradients of the objective (\ssect{sec:method-gradient-optimization}).
These steps can be repeated until a desired number of fiber paths are extracted and optimized.
We then perform a coarse-to-fine optimization by upsampling and optimizing the fiber paths a specified number of times (\ssect{sec:method-coarse-to-fine}).

\subsection{Simulation}
\label{sec:method-simulation}

In this subsection, we describe how we solve the stress field given a shape, some external loads, and a specified fiber layout.
We denote the body as $\Omega$, the stress tensor as $\bm{\sigma}$, the strain tensor as $\bm{\varepsilon}$, the displacement vector as $\mathbf{u}$, and the stiffness tensor as $\texttt{C}$.
The linear elastic model can be written as
\begin{align}
\begin{split}
    -\nabla \cdot \bm{\sigma} & = f, \\
    \bm{\varepsilon} & = \frac{1}{2} \bigl(\nabla \mathbf{u} + (\nabla \mathbf{u})^\intercal\bigr), \\
    \bm{\sigma} & = \texttt{C} : \bm{\varepsilon},
\end{split}
\label{eqn:linear_elasticity}
\end{align}
where the colon is the dot product.
$f$ is the body force and we set it to $0$.
For certain regions on the boundary of $\Omega$ (\ie,~$\partial \Omega$), the value of $\mathbf{u}$ is given as input (\ie, Dirichlet boundary condition).
For the remaining regions, we have $\bm{\sigma} \cdot \mathbf{n} = T$ (\ie, Neumann boundary condition), where $\mathbf{n}$ is the outward unit normal vector, and $T$ is the tractive force which we set to $0$.

The constitutive equations $\bm{\sigma} = \texttt{C} : \bm{\varepsilon}$ can also be written in a matrix product form; under the assumption of in-plane stress, we have
\begin{equation}
    \begin{bmatrix} 
    \sigma_{11} \\
    \sigma_{22} \\
    \sigma_{12} 
    \end{bmatrix}
    =
    \begin{bmatrix}
    \frac{E_1}{1 - \nu_{21}\nu_{12}} & \frac{E_1 \nu_{21}}{1 - \nu_{21} \nu_{12}} & 0 \\
    \frac{E_2 \nu_{12}}{1 - \nu_{12} \nu_{21}} 	& \frac{E_2}{1- \nu_{12} \nu_{21}} & 0 \\
    0 & 0 & \mu
    \end{bmatrix}
    \begin{bmatrix}
    \varepsilon_{11} \\
    \varepsilon_{22} \\
    2\varepsilon_{12} 
    \end{bmatrix},
\end{equation}
where $E_1$ and $E_2$ are Young's moduli, $\nu_{12}$ and $\nu_{21}$ are the Poisson's ratios, and $\mu$ is the shear modulus.
For simplicity, we assume both plastic and fiber are isotropic materials, and they have different Young's moduli $E_{\text{plastic}}$ and $E_{\text{fiber}}$ and identical Poisson's ratio $\nu$.

The next issue is to calculate the Young's modulus field.
Consider a laminate of height $h_{\text{object}}$, with some layers filled with just plastic and others containing both plastic and fiber.  We assume that all layers with fiber, adding up to a total height of $h_{\text{fiber}}$, have identical fiber paths, and omit plastic where fiber is present.
The set of fiber paths is denoted as $P$, and every path $p$ in it is a sequence of vertices on the fiber path.
For a point $x \in \Omega$, for the purpose of differentiability, we define its ``soft'' Young's modulus as
\begin{equation}
    E(x) \coloneqq E_{\text{plastic}} \cdot \alpha_{\text{plastic}}(x) + E_{\text{fiber}} \cdot \alpha_{\text{fiber}}(x),
\end{equation}
where
\begin{equation}
    \alpha_{\text{fiber}}(x) \coloneqq \sum_{p \in P} \exp \left( -\left( \frac{\text{dis}(p, x)}{w_{\text{fiber}} / 2} \right)^2 \right) \cdot h_{\text{fiber}},
\end{equation}
where $w_{\text{fiber}} = 0.9$ mm is the width of the fiber, $\text{dis}(\cdot, \cdot)$ measures the distance between a point and a path, and
\begin{equation}
    \alpha_{\text{plastic}}(x) \coloneqq h_{\text{object}} - \min(\alpha_{\text{fiber}}(x), h_{\text{fiber}}).
\end{equation}
We allow fiber paths to overlap in this setting, as even in real prints from the Markforged Mark Two, we do not observe any problems. We then have ${\mu(x) = \frac{E(x)}{2(1 + \nu)}}$. Finally, we solve the PDE in~\eqn{eqn:linear_elasticity} using FEniCS with DOLFIN \citep{logg2010dolfin} by solving its first-order condition on linear triangular finite elements.  \fig{fig:pipeline} visualizes an example of the calculated stress field, using line integral convolution~\cite{Cabral93}.

\subsection{Greedy fiber path extraction}
\label{sec:method-greedy}

In this subsection, we describe how we greedily extract a fiber path from a stress field along the directions of maximum tensile stress or perpendicular to the direction of maximum compressive stress.
With the stress tensor $\bm{\sigma}$ calculated from \ssect{sec:method-simulation}, we first calculate the stress on plastic:
\begin{equation}
    \bm{\sigma}_{\text{plastic}} \coloneqq \bm{\sigma} \cdot \frac{E_{\text{plastic}} \cdot \alpha_{\text{plastic}}}{E_{\text{plastic}} \cdot \alpha_{\text{plastic}} + E_{\text{fiber}} \cdot \alpha_{\text{fiber}}}.
\end{equation}
For the stress matrix at any point $x \in \Omega$, we can calculate its eigenvalue with the largest absolute value $\lambda(x)$ and the corresponding eigenvector $\mathbf{v}(x)$.
We then build a scalar field with $|\lambda(x)|$ and randomly sample a starting point $x_0$ with the field as sampling weights.
From the starting point, we walk in both directions along with $\pm \mathbf{v}(x_0)$ (or perpendicular to $\mathbf{v}(x_0)$ if~$\lambda(x_0)$ is negative) at a fixed step size of 0.5 mm.
If we walk outside $\Omega$ or within 1.3 mm to $\partial \Omega$ (number measured from prints from Eiger), we retry at most 19 times with a random rotation uniformly sampled between $-\pi / 12$ to $\pi / 12$.
The algorithm stops when a preset length limit is reached, or we cannot walk in both directions even after retries.

We then downsample the extracted fiber path by keeping 1 of every 20 vertices.
We iterate every subsequence of the downsampled path and select the one that minimizes the objective function we will define in~\sect{sec:method-gradient-optimization}.
We repeat this process 10 times (sampling 10 starting points) and keep the one that minimizes the objective function.

\subsection{Gradient calculation and optimization}
\label{sec:method-gradient-optimization}

In this subsection, we describe how we design an objective function and minimize it using a gradient-based optimizer.
We denote the optimized strain energy in~\eqn{eqn:linear_elasticity} as~$U$, and the set of fiber paths as~$P$.
The objective $\mathcal{L}(P)$ is defined as
\begin{equation}
    \small
    -U + \sum_{p \in P} \left( w_{\text{lap}} \cdot \mathcal{L}_{\text{lap}}(p) + w_{\text{min\_l}} \cdot \mathcal{L}_{\text{min\_l}}(p) + w_{\text{bdy}} \cdot \mathcal{L}_{\text{bdy}}(p) \right),
\end{equation}
where $w_{\text{lap}}$, $w_{\text{min\_len}}$, and $w_{\text{bdy}}$ are hyper-parameters.
The Laplacian regularizer $\mathcal{L}_{\text{lap}}$ penalizes non-smooth fiber paths:
\begin{equation}
    \mathcal{L}_{\text{lap}}(p) \coloneqq s(P)^3 \cdot \sum_{i=2}^{|p| - 1} \left| \left| p_i - \frac{p_{i - 1} + p_{i + 1}}{2} \right| \right|^2\, ,
\end{equation}
where $s(P)$ is a count of the total number of segments in all fiber paths (\ie, ${\sum_{p \in P} |p| - |P|}$).
The reason to apply the $s(P)^3$ multiplier is because the Laplacian regularizer is sensitive to upsampling, which we discuss in~\ssect{sec:method-coarse-to-fine}, and this multiplier keeps our Laplacian regularizer scale-invariant.
The minimum-length regularizer $\mathcal{L}_{\text{min\_l}}$ penalizes infeasibly-short fiber paths:
\begin{equation}
    \mathcal{L}_{\text{min\_l}}(p) \coloneqq \max \left( l_\text{min} -  \sum_{i=2}^{|p|} \left| \left| p_i - p_{i - 1} \right| \right|, 0 \right)^2,
\end{equation}
where $l_\text{min}$ is the minimum fiber length that can be printed by the 3D printer.
The boundary regularizer $\mathcal{L}_{\text{bdy}}$ penalizes fiber paths outside $\Omega$ or too close to $\partial \Omega$:
\begin{equation}
    \mathcal{L}_{\text{bdy}}(p) \coloneqq \sum_i \max(d_\text{min} - \text{dis}(p_i, \Omega), 0)^2,
\end{equation}
where $\text{dis}(p_i, \Omega)$ measures the distance from $p$ to $\partial \Omega$ (positive for $p_i \in \Omega$, negative otherwise) and $d_\text{min}$ is the lower limit of distance from fiber to the boundary.

The next step is to calculate $\frac{\mathrm{d} \mathcal{L}(P)}{\mathrm{d} P}$.
Here we apply the adjoint method.
Denote the first-order condition of \eqn{eqn:linear_elasticity} as ${F(\mathbf{u}, P) = 0}$.
By the implicit function theorem (under proper regularity conditions) $\mathbf{u}$ can be thought of a function of $P$, and the derivative $\frac{\mathrm{d} \mathbf{u}}{\mathrm{d} P}$ is well-defined.
Taking the derivative of $F$ with respect to $P$, we have
\begin{equation}
    \frac{\mathrm{d} F}{\mathrm{d} P} = \frac{\partial F}{\partial \mathbf{u}} \frac{\mathrm{d} \mathbf{u}}{\mathrm{d} P} + \frac{\partial F}{\partial P} = 0,
\end{equation}
which leads to
\begin{equation}
    \frac{\mathrm{d} \mathcal{L}(P)}{\mathrm{d} P} = - \frac{\partial \mathcal{L}(P)}{\partial \mathbf{u}}\left( \frac{\partial F}{\partial \mathbf{u}} \right)^{-1} \frac{\partial F}{\partial P} + \frac{\partial \mathcal{L}(P)}{\partial P}.
\end{equation}
We implement this end-to-end differentiation automatically using dolfin-adjoint~\citep{mitusch2019dolfin} and PyTorch~\citep{NEURIPS2019_9015}.
We use the BFGS implementation in SciPy~\citep{virtanen2020scipy} to minimize $\mathcal{L}(P)$, and again we iterate every subsequence of the optimized path and select the one that minimizes $\mathcal{L}(P)$. We can repeat the steps in \ssect{sec:method-simulation}, \ssect{sec:method-greedy}, and \ssect{sec:method-gradient-optimization} several times to extract multiple fiber paths.

\subsection{Coarse-to-fine optimization}
\label{sec:method-coarse-to-fine}

To speed up the optimization, we perform multigrid optimization.
As described in \ssect{sec:method-greedy}, we initially downsample all the fiber paths.
Then, for every fiber path $p$, we insert midpoints between every $p_i$ and $p_{i + 1}$ by B-spline interpolation, using SciPy, and optimize all the fiber paths.
This process can be repeated several times to generate the final fiber paths for 3D printing.

%% file: figText/pipeline.tex
\begin{figure*}[t]
    \centering
    \includegraphics[width=\linewidth]{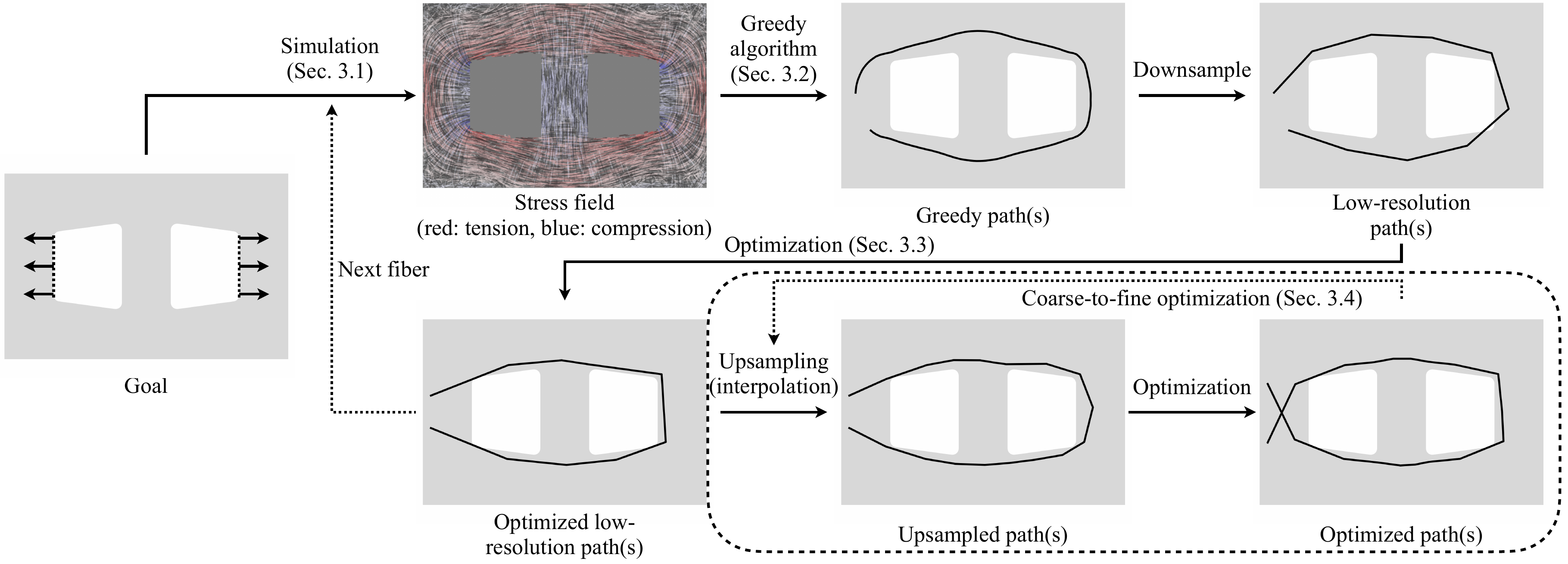}
    \caption{We repeatedly use the finite element method to calculate the stress field of the object (\ssect{sec:method-simulation}), extract a new fiber path by greedily ``walking'' on the stress field (\ssect{sec:method-greedy}), optimize the downsampled fiber path with an objective function designed to maximize stiffness and regularize fiber paths to be manufacturable (\ssect{sec:method-gradient-optimization}), and finally upsample and optimize all the fiber paths several times to perform coarse-to-fine optimization (\ssect{sec:method-coarse-to-fine}).}
    \label{fig:pipeline}
\end{figure*}

%% file: text/hardware.tex
\section{Fabrication and Experimental setup}

\label{sec:hardware}

\input{figText/instron}

In this section, we describe how we manufacture real 3D prints and measure their position-load curves.
We use a Markforged Mark Two printer with nylon as the plastic material and carbon fiber as the reinforcing fiber material.
We print laminates with a height of 2 mm and 16 layers, of which the 4th, 7th, 10th, and 13th layers are fiber layers.
All layers without fiber and regions in fiber layers without fiber are filled with nylon (solid fill).

For the 2D shape, we use a 46 mm $\times$ 30 mm rectangle with two rounded isosceles trapezoidal holes, the same shape as shown in~\fig{fig:pipeline}.
The isosceles trapezoids have two sides of 11 mm and 14 mm and a height of 11 mm, with every corner smoothed by an arc with a radius of 1 mm.
We will reuse this shape in~\ssect{sec:modulus},~\ssect{ssec:case-3}, and~\ssect{sec:ablation}.

To measure the position-load curve of a print, we insert two square nuts into both its holes and apply tension to them using a universal testing machine (Instron 600DX), as shown in~\fig{fig:instron}.
The machine is programmed to move at a speed of 20 mm/min until the object breaks or by a manual stop when we believe enough data is collected. 
A position-load curve is recorded for every print.

%% file: figText/instron.tex
\begin{figure}[t]
    \centering
    \includegraphics[width=0.85\linewidth]{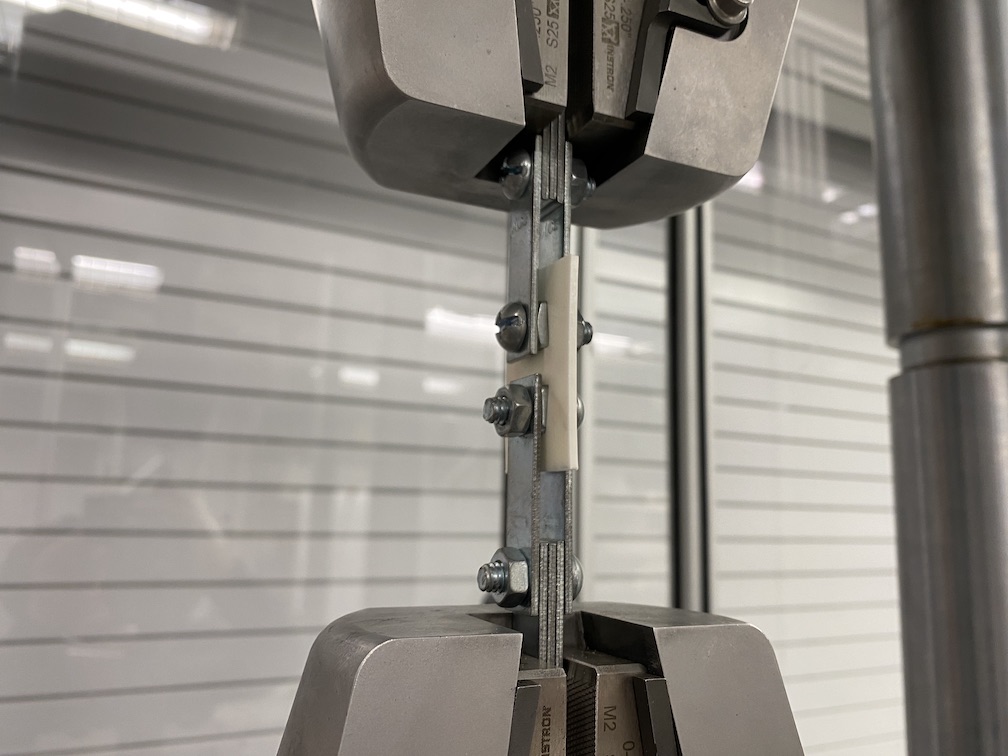}
    \caption{A 3D printed part being tested on a universal testing machine (Instron 600DX), with square nuts in the trapezoidal holes. The machine moves at a speed of 20 mm/min and stops when the object breaks or by a manual stop.}
    \label{fig:instron}
\end{figure}

%% file: text/modulus.tex
\section{Modulus calculation}
\label{sec:modulus}

In this section, we describe how we determine the effective Young's moduli of nylon and carbon fiber.  We print composites with different (baseline) fiber layouts, measure their stiffness, then optimize for moduli such that their stiffness in simulation best matches the real-world measurements.

\subsection{3D prints for testing}

\input{figText/modulus_prints}

We use Eiger to generate nine different layouts of carbon fiber paths: no fiber path, 1 to 3 inner rings, 1 to 3 outer rings, 1 to 2 rings for all walls.
To reduce the bias introduced by the non-uniformity of the material, we print all of them in one batch, as shown in~\fig{fig:modulus_prints}.
Due to the variability of the printing process, we print three batches of these nine prints and pick the batch with the best printing quality.

\subsection{Stiffness measurement}

\input{figText/modulus_pl_curve}

As described in~\sect{sec:hardware}, we test the prints and record their position-load curves (\fig{fig:modulus_pl_curve}).
Note that the beginning of every curve can be noisy as the part is not perfectly vertical, \etc.
Additionally, a large load can cause the part to buckle out of the 2D plane, which violates our in-plane stress assumption.
We therefore measure the position change between a load of 150 N and a load of 300 N for every print, and calculate the stiffness by dividing load change (150~N) by position change (in mm).
The results are shown in~\fig{fig:modulus_results}, marked as ``X''.
Note that there is a factor of 0.5 when converting stiffness in N/mm to energy in N$\cdot$mm at 1 mm displacement (\eg, a stiffness of 500 N/mm corresponds to having strain energy of 250 N$\cdot$mm at 1~mm displacement).

\subsection{Simulation and modulus calculation}

\input{figText/modulus_results}

For each measured data point, we apply Dirichlet boundary conditions corresponding to 1~mm displacement on the two shorter sides of the two holes on the rectangle. We calculate the strain energy of the object, then do a grid search for the values of the moduli of nylon and carbon that minimize the sum of squared distances between measured and simulated stiffness.
The search yields moduli of 0.40 GPa for nylon and 20.1 GPa for carbon, with results shown in~\fig{fig:modulus_results}.
As we can see, the simulation results mostly match the real results, with small residuals relative to the energy.

%% file: figText/modulus_prints.tex
\begin{figure}[t]
    \centering
    \includegraphics[width=\linewidth]{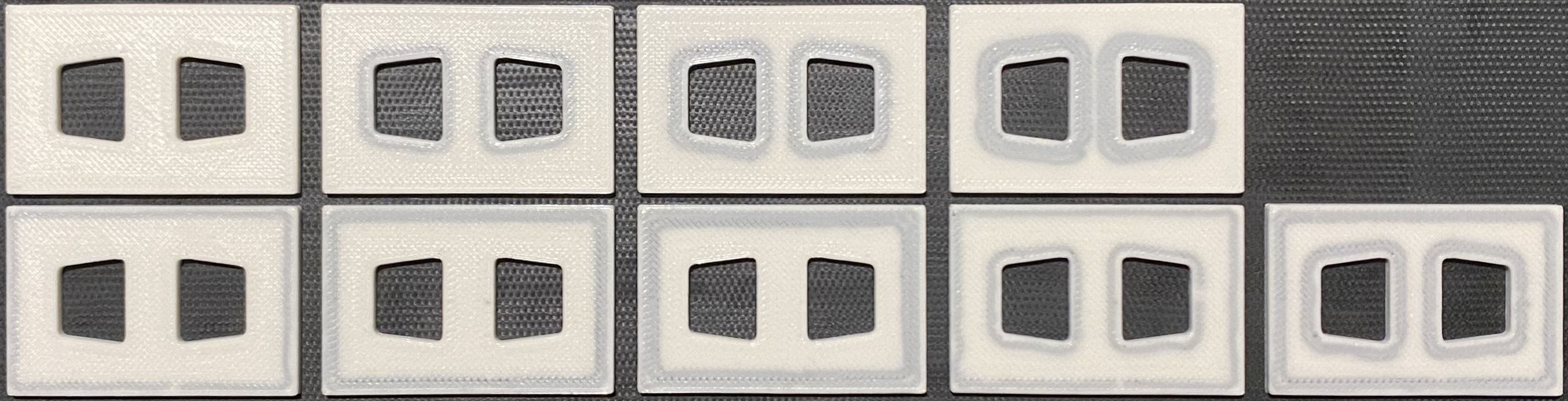}
    \caption{Nine different fiber layouts printed for moduli calculation (left to right, top to bottom): no fiber path, 1 to 3 inner rings, 1 to 3 outer rings, 1 to 2 rings for all walls.}
    \label{fig:modulus_prints}
\end{figure}

%% file: figText/modulus_pl_curve.tex
\begin{figure}[t]
    \centering
    \includegraphics[width=\linewidth]{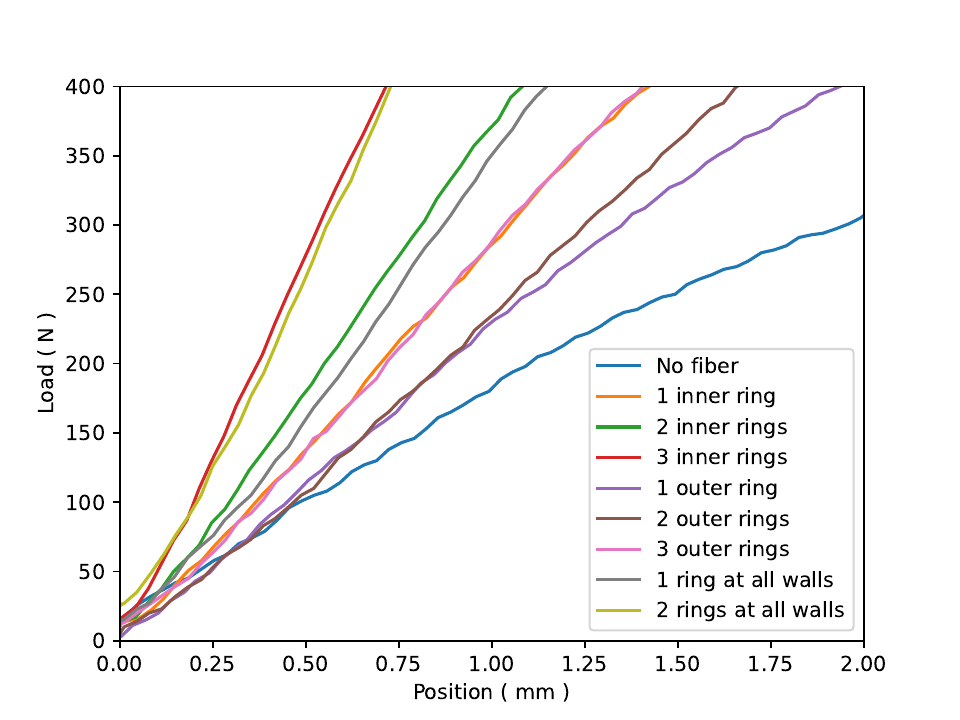}
    \caption{Position-load curves recorded from the testing machine. The beginning parts of the curves are noisy due to parts not being perfectly vertical, \etc, and a too-large load can cause the object to buckle, violating our in-plane stress assumption. We therefore use the middle parts of the curves, with loads between 150 N and 300 N, to calculate the stiffness.}
    \label{fig:modulus_pl_curve}
\end{figure}

%% file: figText/modulus_results.tex
\begin{figure}[t]
    \centering
    \includegraphics[width=\linewidth]{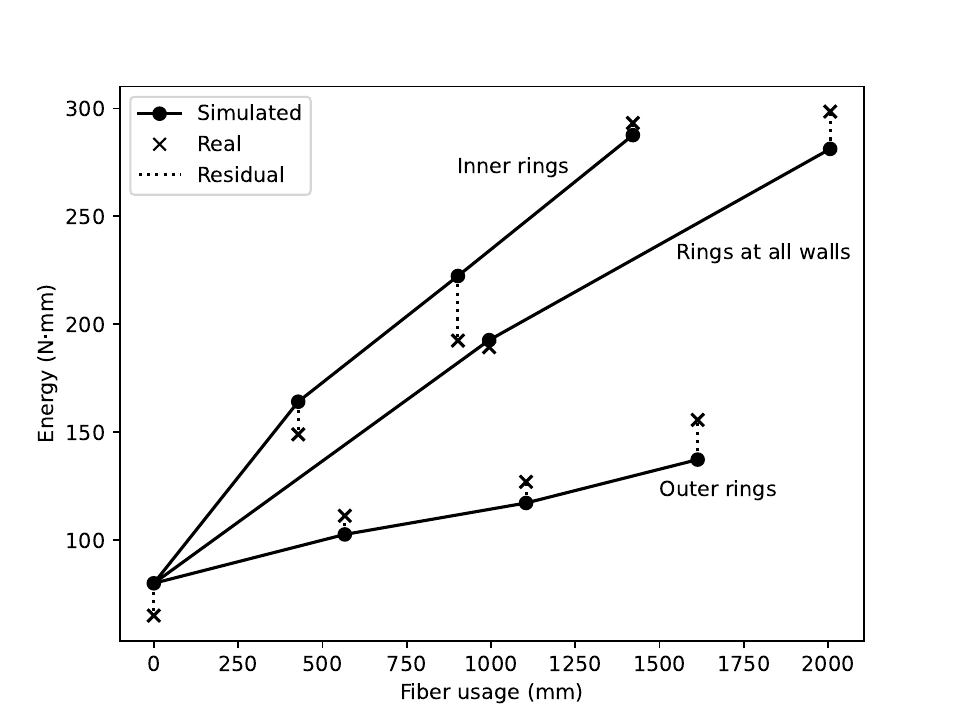}
    \caption{We calibrate the moduli of nylon and carbon fiber using nine prints with three different types of fiber layouts: inner rings, outer rings, and rings at all walls.  The solid lines connect datapoints sharing the same fiber layout strategy.  The real results are marked as ``X'', the simulated results are marked as small solid circles, and the residuals are shown as dotted lines. The energy numbers are calculated at 1 mm displacement.}
    \label{fig:modulus_results}
\end{figure}

%% file: text/experiments.tex
\section{Experiments}
\label{sec:experiments}

\input{figText/shapes}

In this section, we present detailed evaluations of the performance of our method in both simulation and real experiments on four case studies (\ssect{ssec:case-1}--\ssect{ssec:case-4}), then show several additional results in~\ssect{ssec:addition-shapes}.
We start with two simple shapes---a rectangle and a ``plus'' shape---then move to more complex shapes: rectangles with two and four holes (\fig{fig:shapes}).
The first baseline we use is \textit{concentric} fiber rings from Eiger, which have three different types: \textit{inner}, \textit{outer}, and \textit{all walls}.
For the next two case studies, to better illustrate the effectiveness of our algorithm on complex shapes and loads, we include two additional baselines: (1) \textit{greedy}, simplifying our algorithm by removing all the optimization components and directly generating results using the greedy algorithm; (2) \textit{field-opt-greedy}, similar to \textit{greedy} but with an additional step of field optimization before running the greedy algorithm.
The latter baseline, intended to represent the approach of previous work on fiber orientation optimization (see \ssect{sec:prev_orientation}), optimizes a vector field that aligns to the stress direction, with a smoothing regularizer.
Additional details about the field optimization can be found in the appendix.
We refer to the results from our method as \textit{optimized}.
For all the experiments (unless otherwise specified), we use the BFGS optimizer and limit the maximum number of iterations to 500 and a gradient tolerance of $3 \times 10^{-9}$.
The objective function is set with $w_{\text{lap}} = 1 \times 10^{-8}$, $w_{\text{min\_l}} = 1$, and $w_{\text{bdy}} = 1$.

\subsection{Case 1: rectangle}
\label{ssec:case-1}
\input{figText/rectangle_steps}
\input{figText/rectangle_fiber_paths}

In this case study, we show how our algorithm works step by step on a rectangle (45 mm $\times$ 30 mm), with tension applied to its two shorter sides (\fig{fig:rectangle_shape}).
As we can see in the first row in~\fig{fig:rectangle_steps}, we first greedily extract a fiber path and optimize it.
When we add and optimize a second fiber path, the two paths separate and curve.
In the second row, we extract a third greedy path and optimize the three paths together.
Finally, we double the number of points of both fiber paths and optimize the three paths together.
The last step does not help much since the task is relatively simple.
As shown in~\fig{fig:rectangle_fiber_paths}, for a fixed displacement of 1 mm, our algorithm uses less fiber while achieving higher energy in simulation, compared to 1 \textit{outer} concentric fiber ring, which lays fiber in vertical directions that are much less useful than fibers in horizontal directions.

\subsection{Case 2: ``plus'' shape}
\input{figText/plus_fiber_paths}
\input{figText/plus_plot}
\input{figText/rectangle_2_holes_fiber_paths}

As shown in~\fig{fig:plus_shape}, we use a ``plus'' shape whose edges are all of length 15 mm, and we apply tension to two sides of the shape.
We compare fiber paths of \textit{outer} and \textit{optimized} in~\fig{fig:plus_fiber_paths}, with three solutions from each strategy.
As we can see, \textit{outer} lays fibers in regions of low relevance to the loads applied, in contrast to \textit{optimized} which prioritizes regions of high relevance to the loads.
We also observe that the optimization process automatically distributes fiber paths uniformly as we extract more fiber paths.
Based on our simulation, for a fixed displacement at 1 mm, the fiber paths of \textit{optimized} improve upon the Pareto front of \textit{outer}, as shown in~\fig{fig:plus_plot}.

\subsection{Case 3: rectangle with two holes}
\label{ssec:case-3}
\input{figText/rectangle_2_holes_plot_concentric}
\input{figText/rectangle_2_holes_plot_greedy}

As shown in~\fig{fig:rectangle_2_holes_shape}, we also tested a rectangle (46 mm $\times$ 30 mm) with two rounded isosceles trapezoid holes, with external forces applied to the two sides of the holes.

\paragraph{Planned fiber paths and simulation results}
The fiber paths generated from all methods are shown in~\fig{fig:rectangle_2_holes_fiber_paths}.
We set the maximum greedy fiber path length so that fiber lengths of \textit{greedy}, \textit{field-opt-greedy}, and \textit{optimized} are comparable.
As we can see, the baseline methods use only geometric information; both \textit{greedy} and \textit{field-opt-greedy} generate similar fiber paths along stress directions, but paths from \textit{field-opt-greedy} are smoother; \textit{optimized} wraps fiber paths tightly around the holes while aligned with stress direction, yielding larger strain energy when using a similar amount of fiber.

\paragraph{Real experiment results}
To evaluate the quality of fiber paths, we perform real-world experiments by applying tension to 3D prints on a universal testing system (600DX from Instron).
Due to the limited space on the printer bed, two sets of comparisons are performed separately: (1) \textit{inner}, \textit{outer}, and \textit{all walls} \textit{vs.} \textit{optimized}; (2) \textit{greedy} and \textit{field-opt-greedy} \textit{vs.} \textit{optimized}.
We thus printed eight batches, four for each set of comparisons.
Again, as in~\sect{sec:hardware}, we measure the stiffness of a print by calculating the slope of its position-load curve, picking two points that have loads of 150 N and 300 N.
The results of \textit{inner}, \textit{outer}, and \textit{all walls} \textit{vs.} \textit{optimized} are shown in~\fig{fig:rectangle_2_holes_plot_concentric}.
As we can see, our algorithm consistently provides significantly higher stiffness than the concentric baselines when using a similar or lower amount of fiber.
Note that the fiber lengths may have slight discrepancies between simulation and real-world experiments since they are from different path generation algorithms (one from our re-implementation of Eiger, another from Eiger directly).
The results of \textit{greedy} and \textit{field-opt-greedy} \textit{vs.} \textit{optimized} are shown in~\tbl{tbl:rectangle_2_holes_plot_greedy}. Again, our algorithm consistently improves the stiffness over the two baselines while using a similar or lower amount of fiber.

\subsection{Case 4: rectangle with four holes}
\label{ssec:case-4}
\input{figText/rectangle_4_holes_fiber_paths}
\input{figText/rectangle_4_holes_plot}
As shown in~\fig{fig:rectangle_4_holes_shape}, we also tested a rectangle (84 mm $\times$ 28 mm) with four rounded isosceles trapezoid holes.
We design the shape to be multi-functional---if we label the holes from 1 to 4 from left to right, we assume the user uniformly chooses one of the four settings: 1) hole 1 and hole 3; 2) hole 1 and hole 4; 3) hole 2 and hole 3; 4) hole 2 and hole 4.
To support this multi-functional shape, we simulate all four cases and calculate the average strain energy.

The fiber paths from all the methods are shown in~\fig{fig:rectangle_4_holes_fiber_paths}.
Again, both \textit{greedy} and \textit{field-opt-greedy} produce fibers along stress directions with fiber paths from \textit{field-opt-greedy} being slightly smoother.
\textit{Optimized} lays the first fiber over all holes and lays the second fiber around the middle two holes, due to the multi-functional nature of the shape.
The energy-fiber usage plot is shown in~\fig{fig:rectangle_4_holes_plot}, where \textit{optimized} improves upon the Pareto front of every baseline.

\subsection{Results on additional shapes}
\label{ssec:addition-shapes}
\input{figText/additional_shapes}
In this subsection, we provide results from our method and baselines on several additional shapes.
The shape designs are inspired by sketches from SketchGraphs~\citep{seff2020sketchgraphs}, a large-scale dataset of sketches of real-world CAD models, as well as shapes from existing works~\citep{shafighfard2019design, ma2022strength}.
We use a Laplacian regularizer weight $w_{\text{lap}}=5\times10^{-7}$, and the results are shown in~\fig{fig:additional_shapes}, with every dotted line a Dirichlet boundary condition.
For the first shape, all methods use a similar amount of fiber but \textit{optimized} achieves much higher energy than others.
For the second shape, \textit{optimized} uses a similar amount of fiber as \textit{concentric}, less fiber than \textit{greedy} and \textit{field-opt-greedy} but achieves higher energy.
For the third shape, \textit{optimized} achieves comparable energy as \textit{concentric} but saves approximately 70\% of fiber. Compared to \textit{greedy} and \textit{field-opt-greedy}, \textit{optimized} achieves much higher energy while using slightly more fiber.
For the fourth and fifth shapes, \textit{optimized} uses less fiber or comparable fiber as other baselines while achieving significantly higher energy.

%% file: figText/shapes.tex
\begin{figure}[t]
    \centering
    \begin{subfigure}{0.49\linewidth}
        \includegraphics[width=\linewidth]{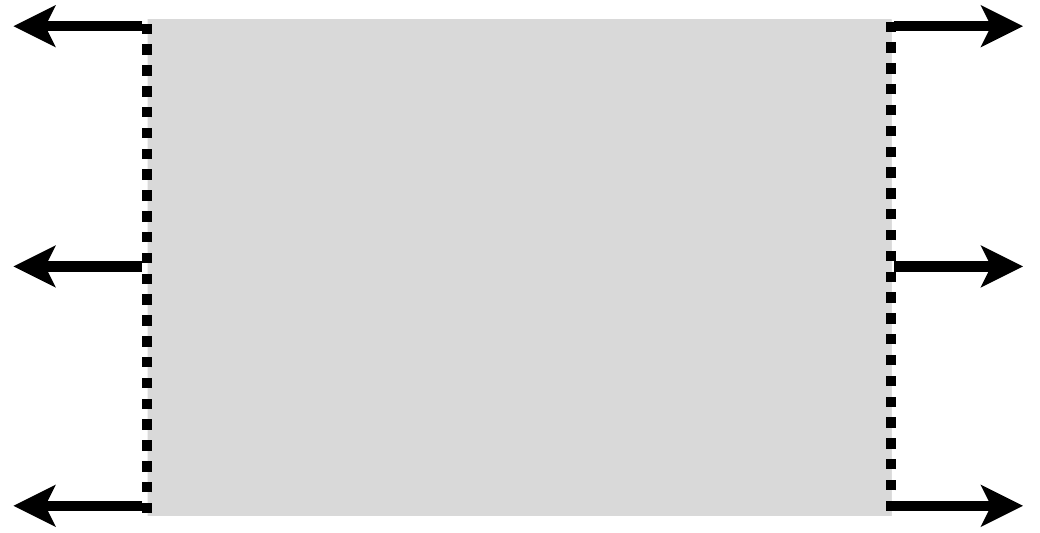}
        \caption{Rectangle}
        \label{fig:rectangle_shape}
    \end{subfigure}
    \begin{subfigure}{0.49\linewidth}
        \includegraphics[width=\linewidth]{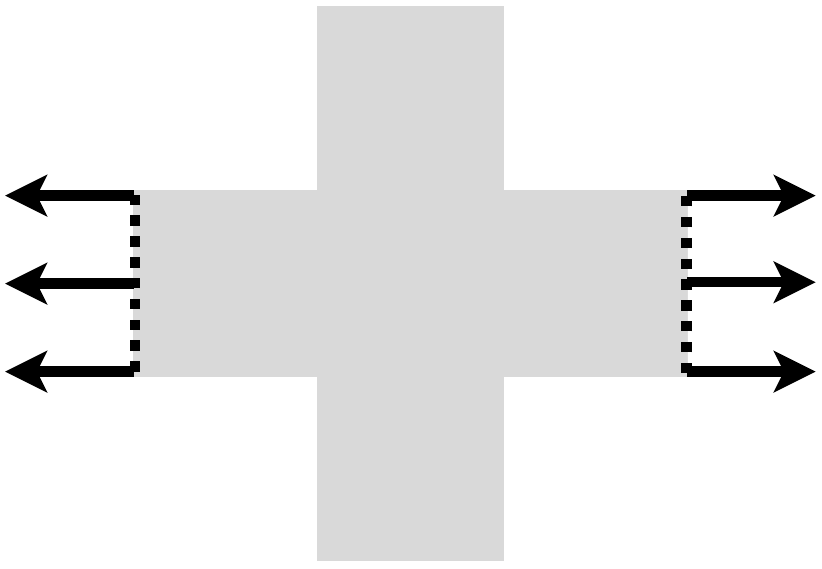}
        \caption{Plus shape}
        \label{fig:plus_shape}
    \end{subfigure}
    \begin{subfigure}{0.49\linewidth}
        \includegraphics[width=\linewidth]{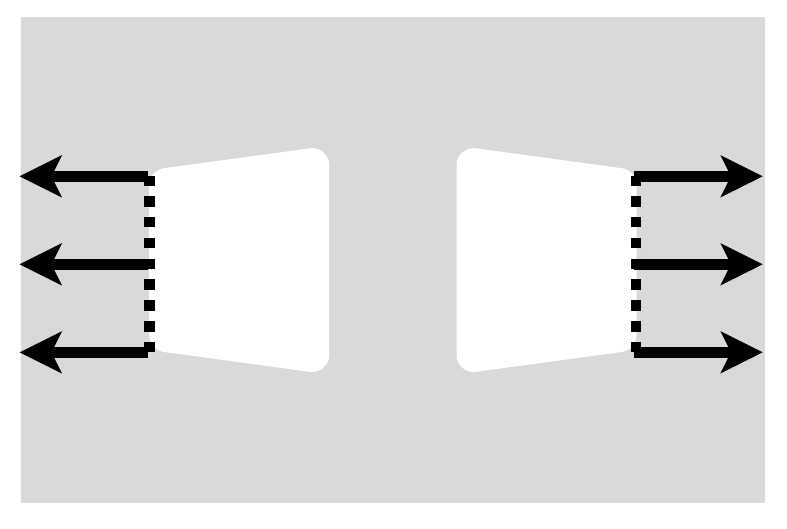}
        \caption{Rectangle with two holes}
        \label{fig:rectangle_2_holes_shape}
    \end{subfigure}
    \begin{subfigure}{0.49\linewidth}
        \includegraphics[width=\linewidth]{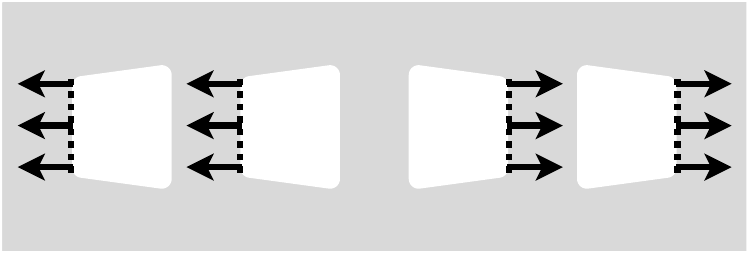}
        \caption{Rectangle with four holes}
        \label{fig:rectangle_4_holes_shape}
    \end{subfigure}
    \caption{Four shapes we use in our case studies. (a) and (b) are relatively simple shapes, and the loads are applied on the two sides. For (c), a rectangle with two holes, tension is applied on the two shorter sides of the holes. (d) is designed to be a multi-functional rectangle with four holes, and the user can choose one hole from the left two holes and another hole from the right two holes to apply tension.}
    \label{fig:shapes}
\end{figure}

%% file: figText/rectangle_steps.tex
\begin{figure}[t]
    \centering
    \includegraphics[width=\linewidth]{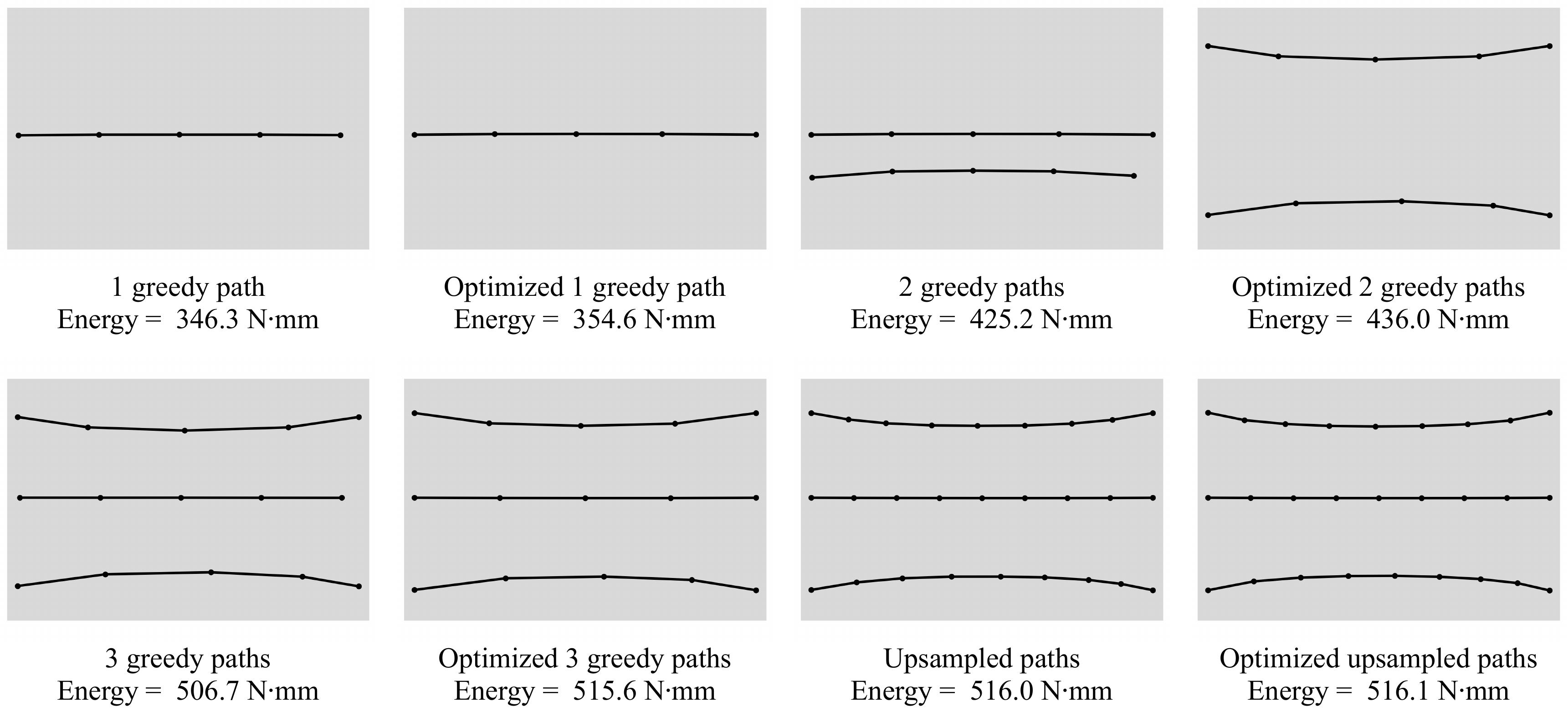}
    \caption{Step-by-step visualization of how our method extracts three fiber paths, optimizes them, and performs coarse-to-fine optimization on the \textit{rectangle} shape. In the first row, we extract the first fiber path, optimize it, extract the second fiber path, and optimize both paths. The two paths curve and move up and down after the optimization, respectively. In the second row, we extract a third fiber path, optimize all three paths, upsample them, and finally optimize them. The energy numbers are calculated at 1 mm displacement.}
    \label{fig:rectangle_steps}
\end{figure}

%% file: figText/rectangle_fiber_paths.tex
\begin{figure}[t]
    \centering
    \includegraphics[width=0.8\linewidth]{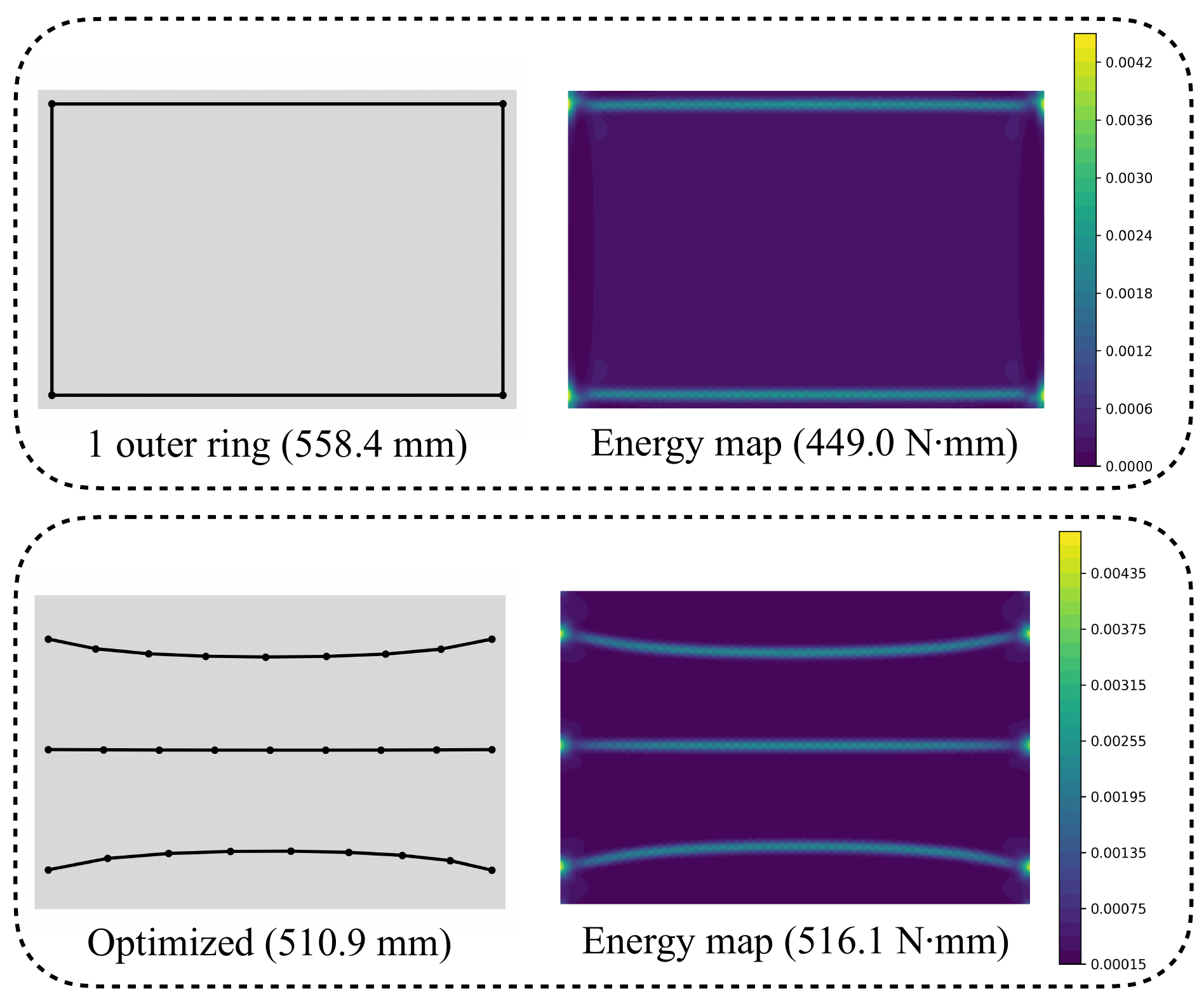}
    \caption{Fiber paths and energy maps of \textit{outer} and \textit{optimized} at 1 mm displacement on the \textit{rectangle} shape. We use less fiber while achieving higher energy, as the baseline lays vertical fibers that are much less useful than horizontal fibers.}
    \label{fig:rectangle_fiber_paths}
\end{figure}

%% file: figText/plus_fiber_paths.tex
\begin{figure}[t]
    \centering
    \includegraphics[width=\linewidth]{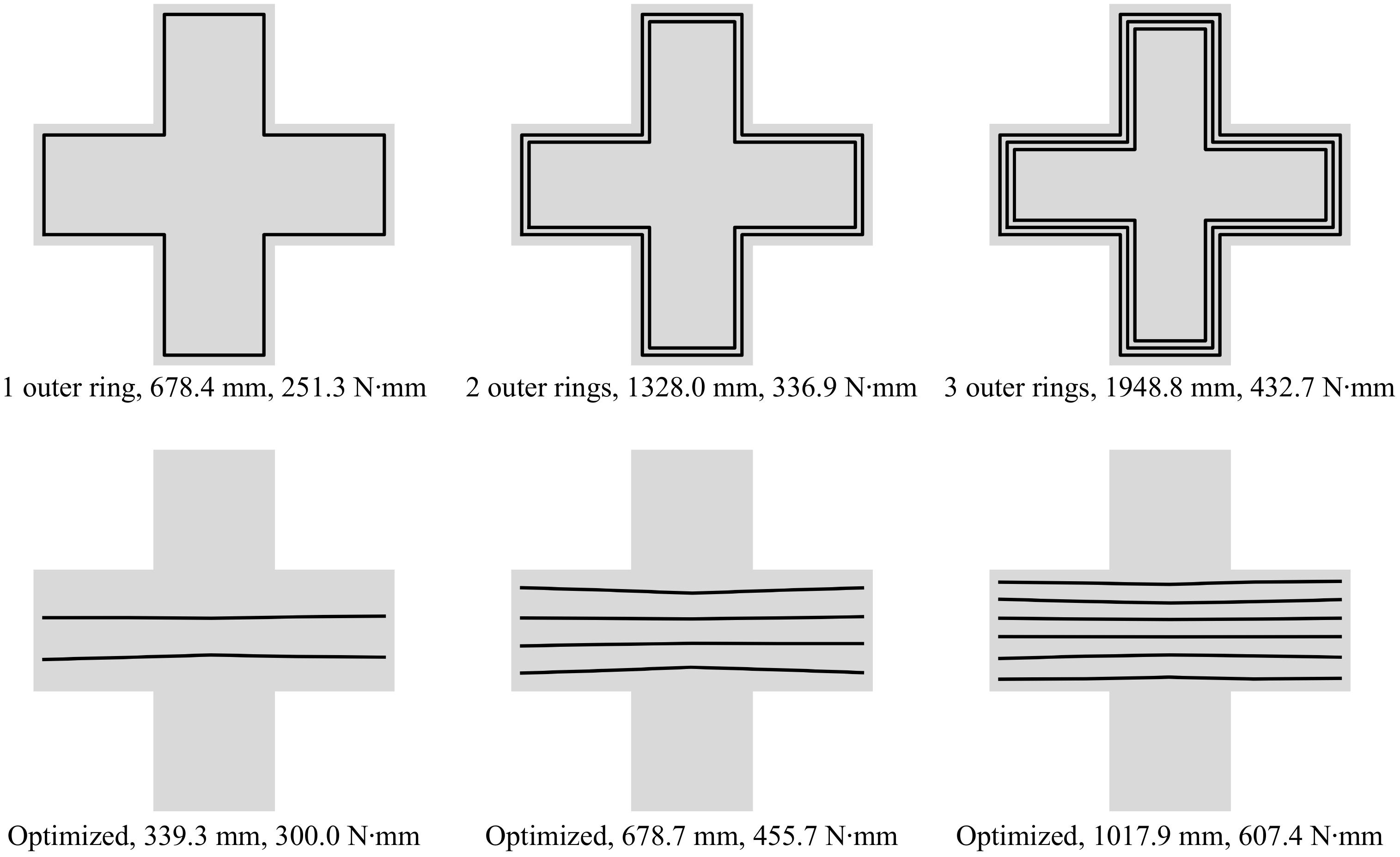}
    \caption{Fiber paths, lengths, and strain energy at 1 mm displacement of \textit{outer} and \textit{optimized} on the \textit{plus} shape. With the help of optimization, fiber paths automatically distribute themselves uniformly in the space as we increase the number of fiber paths. By laying slightly bending fibers in horizontal directions, we save fiber while increasing the energy, compared to \textit{outer}, which lays fiber in unrelated regions.}
    \label{fig:plus_fiber_paths}
\end{figure}

%% file: figText/plus_plot.tex
\begin{figure}[t]
    \centering
    \includegraphics[width=0.8\linewidth]{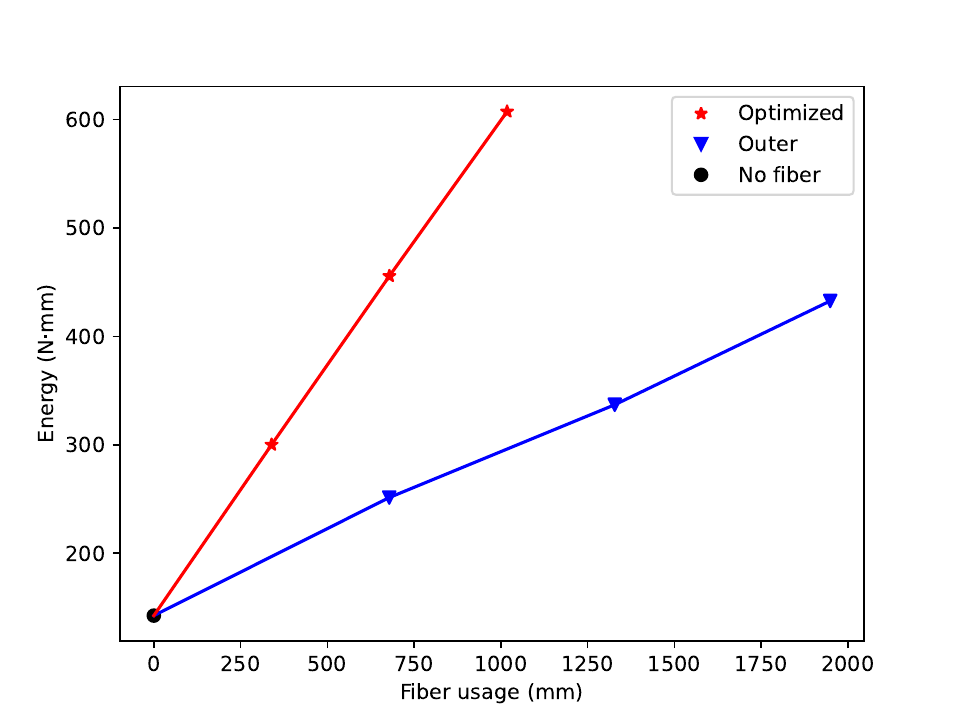}
    \caption{Energy-fiber usage plot of \textit{outer} and \textit{optimized} at 1 mm displacement on the \textit{plus} shape. Our method improves over the Pareto front of \textit{outer} by laying fibers according to the external loads.}
    \label{fig:plus_plot}
\end{figure}

%% file: figText/rectangle_2_holes_fiber_paths.tex
\begin{figure*}[t]
    \centering
    \includegraphics[width=\linewidth]{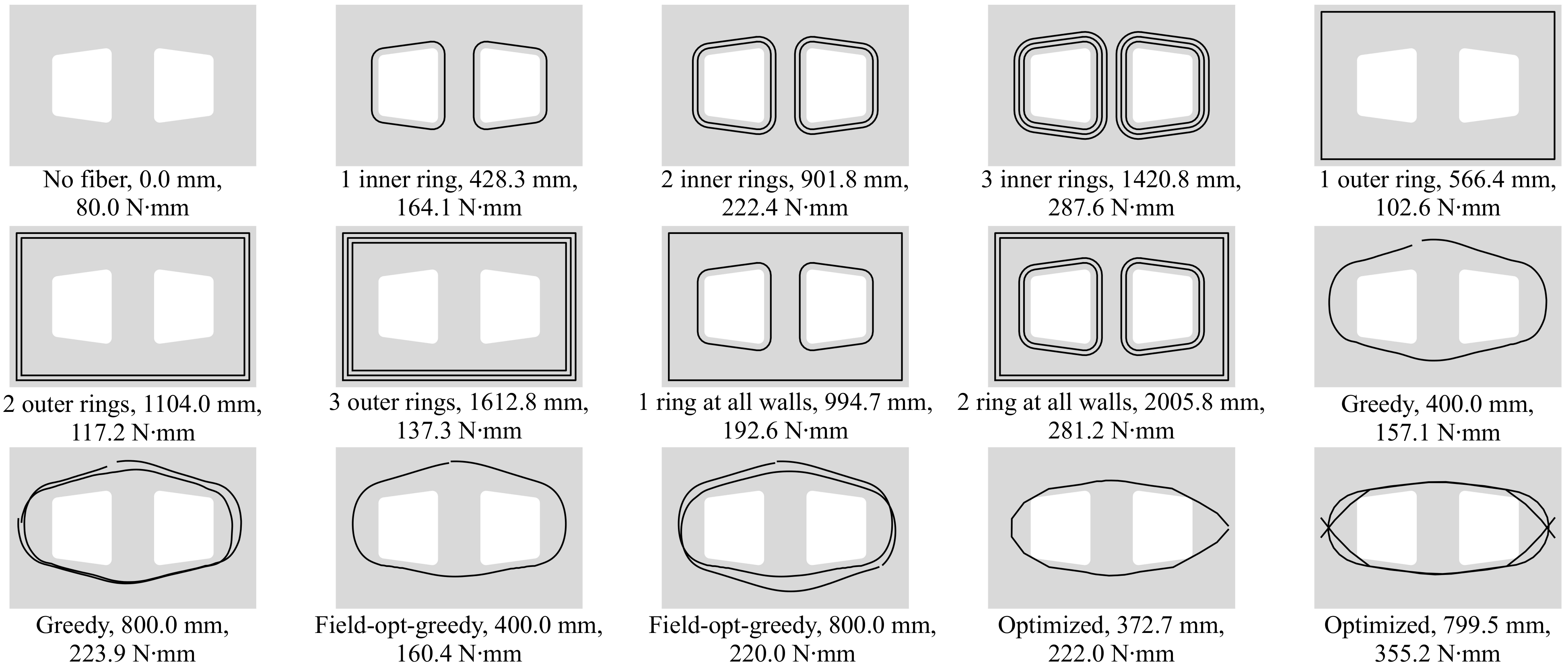}
    \caption{Fiber paths, lengths, and strain energy at 1 mm displacement of \textit{inner}, \textit{outer}, \textit{all walls}, \textit{greedy}, \textit{field-opt-greedy}, and \textit{optimized} on the \textit{rectangle with two holes} shape. \textit{field-opt-greedy} provides similar but smoother paths compared to \textit{greedy}, and \textit{optimized} provides more effective fiber paths. Note that there is a factor of 2 when converting the strain energy in N·mm at 1 mm displacement to stiffness in N/mm which we will use in real experiments (\eg, strain energy of 250 N$\cdot$mm at 1 mm displacement corresponds to having a stiffness of 500 N/mm).}
    \label{fig:rectangle_2_holes_fiber_paths}
\end{figure*}

%% file: figText/rectangle_2_holes_plot_concentric.tex
\begin{figure*}[t]
    \centering
    \begin{subfigure}{0.24\linewidth}
        \includegraphics[width=\linewidth]{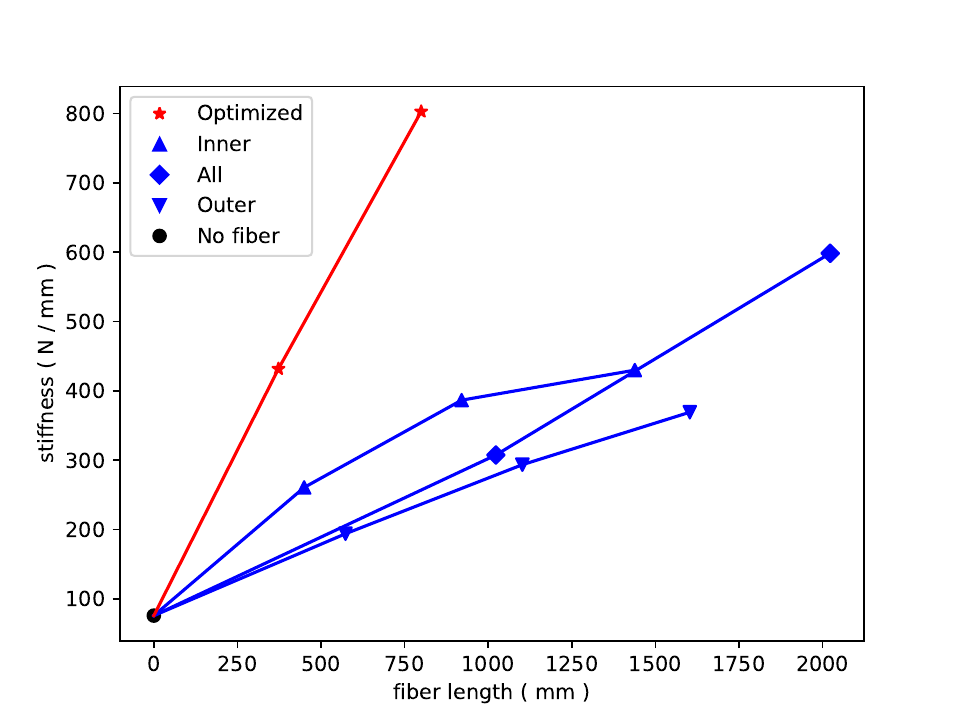}
        \caption{Batch 1}
    \end{subfigure}
    \begin{subfigure}{0.24\linewidth}
        \includegraphics[width=\linewidth]{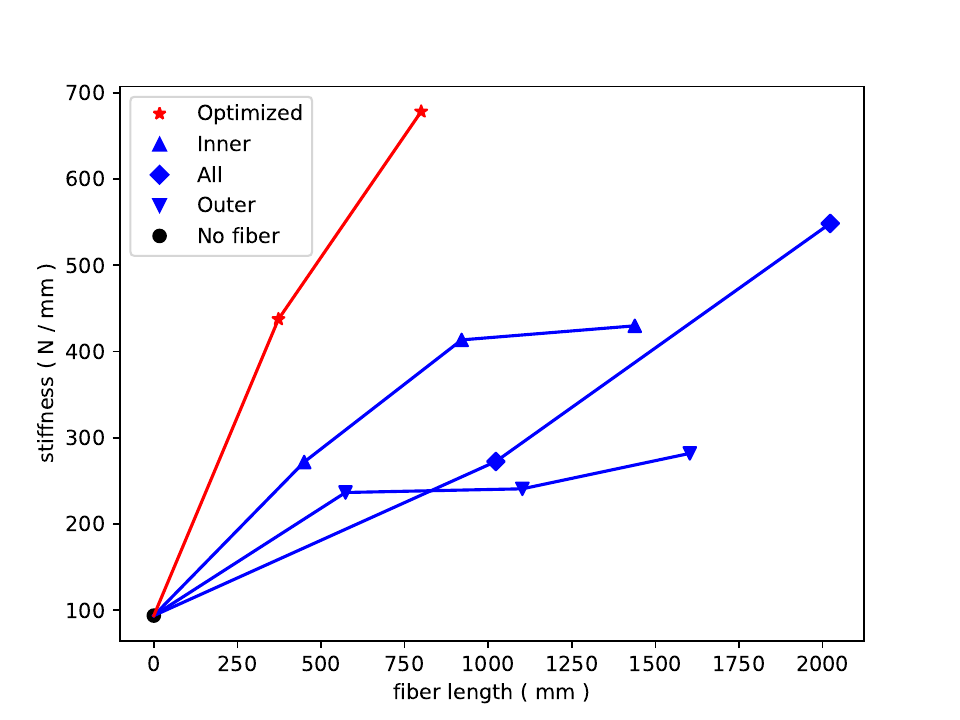}
        \caption{Batch 2}
    \end{subfigure}
    \begin{subfigure}{0.24\linewidth}
        \includegraphics[width=\linewidth]{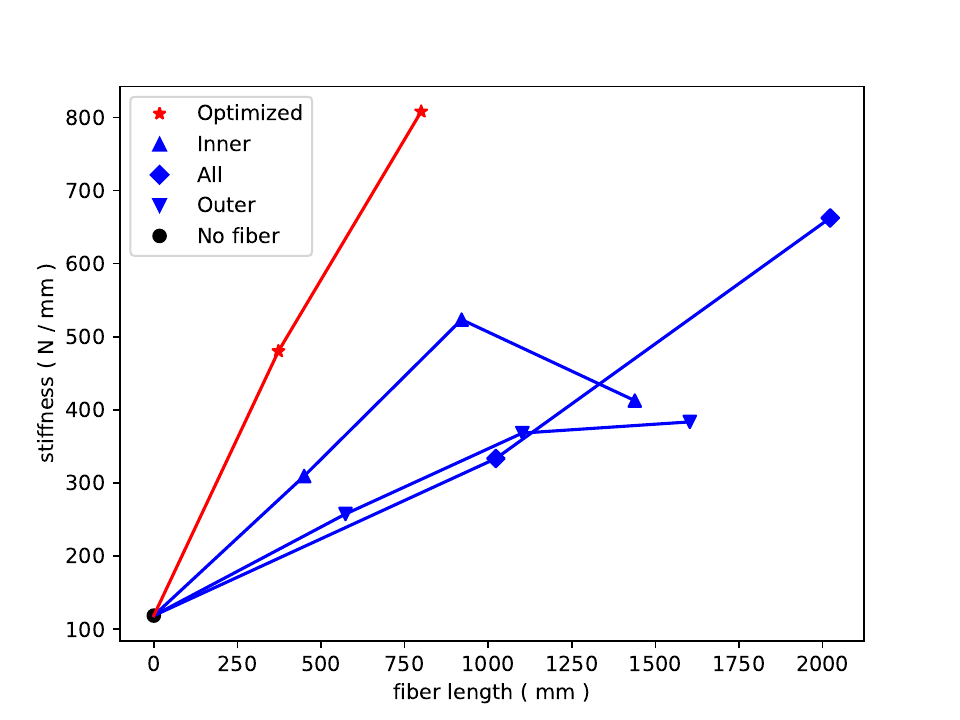}
        \caption{Batch 3}
    \end{subfigure}
    \begin{subfigure}{0.24\linewidth}
        \includegraphics[width=\linewidth]{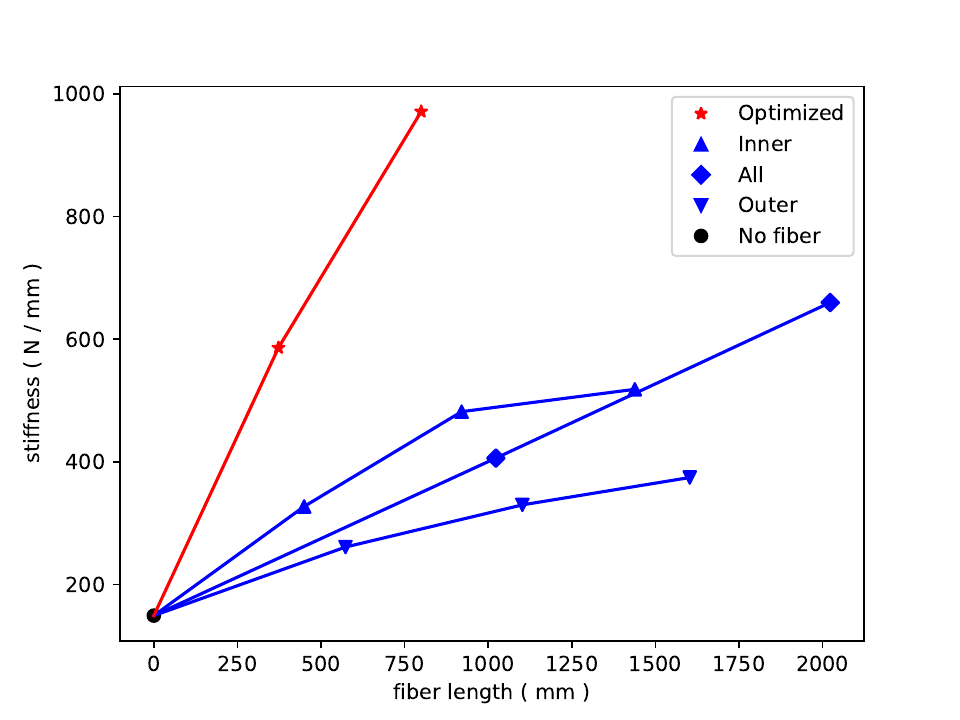}
        \caption{Batch 4}
    \end{subfigure}
    \caption{\textbf{Real} stiffness-fiber length plots of \textit{inner}, \textit{outer}, \textit{all walls}, and \textit{optimized} on the \textit{rectangle with two holes} shape, measured between 150 N and 300 N (4 batches). By laying fibers tightly around the holes, \textit{optimized} consistently performs better than all others.}
    \label{fig:rectangle_2_holes_plot_concentric}
\end{figure*}

%% file: figText/rectangle_2_holes_plot_greedy.tex
\begin{table}[t]
    \setlength{\tabcolsep}{4pt}
    \caption{\textbf{Real} (measured) stiffness of \textit{greedy} (\textit{g}), \textit{field-opt-greedy} (\textit{f}), and \textit{optimized} (\textit{o}) on the \textit{rectangle with two holes} shape, measured between 150 N and 300 N (4 batches). \textit{Optimized} performs consistently better than the baselines when using a similar or less amount of fiber.}
 	\centering
    \begin{tabular}{ccccccc}
    \toprule
    \multirow{2}{*}{Stiffness (N/mm)} & \multicolumn{3}{c}{Solution 1} & \multicolumn{3}{c}{Solution 2} \\
    \cmidrule(lr){2-4}\cmidrule(lr){5-7}
    & \textit{g} & \textit{f} & \textit{o} & \textit{g} & \textit{f} & \textit{o} \\
    \midrule
    Batch 1 & 490.1 & 574.6 & 625.5 & 745.0 & 741.3 & 992.6 \\
    Batch 2 & 584.3 & 692.0 & 756.0 & 801.0 & 741.3 & 985.2 \\
    Batch 3 & 483.5 & 485.0 & 656.7 & 671.6 & 603.4 & 953.5 \\
    Batch 4 & 491.7 & 481.7 & 603.0 & 670.0 & 670.9 & 970.4 \\
    \midrule
    Average & 512.4 & 558.3 & \textbf{660.3} & 721.9 & 689.2 & \textbf{975.4} \\
    Length (mm) & 400.0 & 400.0 & 372.7 & 800.0 & 800.0 & 799.5 \\
    \bottomrule
    \end{tabular}
    \label{tbl:rectangle_2_holes_plot_greedy}
\end{table}

%% file: figText/rectangle_4_holes_fiber_paths.tex
\begin{figure*}[t]
    \centering
    \includegraphics[width=\linewidth]{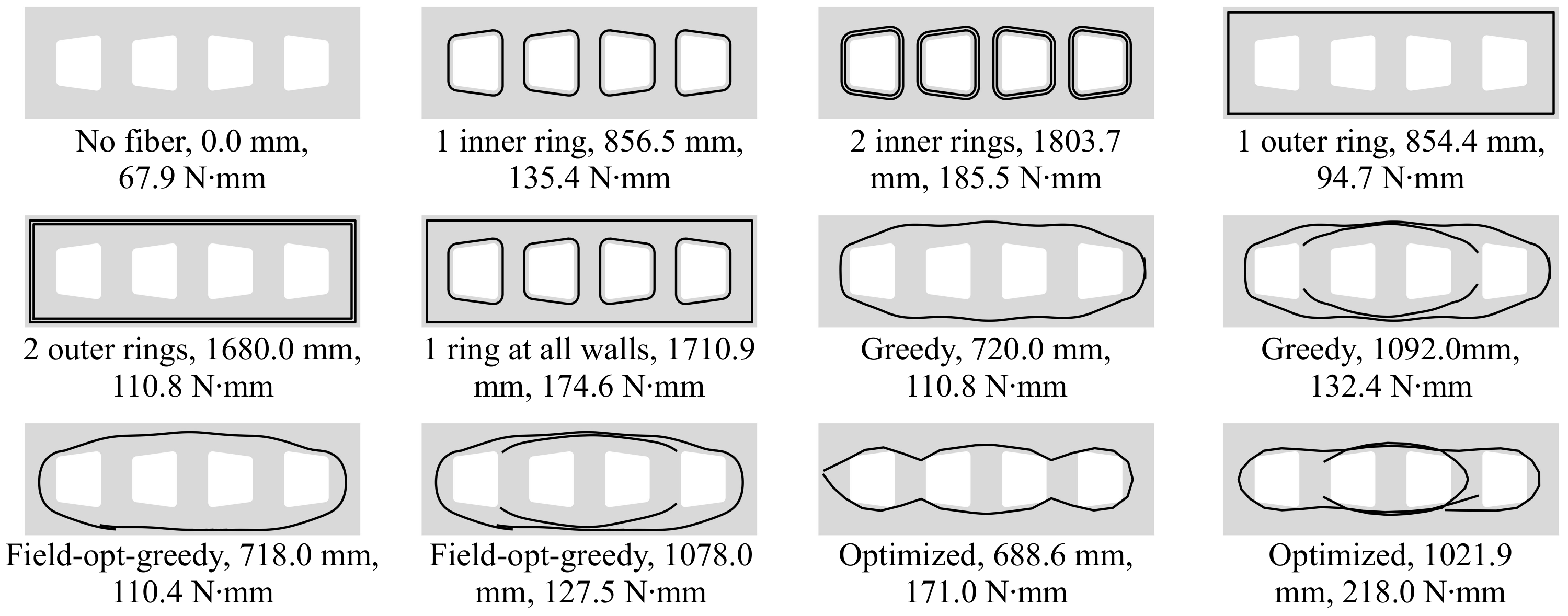}
    \caption{Fiber paths, lengths, and strain energy at 1 mm displacement of \textit{inner}, \textit{outer}, \textit{all walls}, \textit{greedy}, \textit{field-opt-greedy}, and \textit{optimized} on the multi-functional \textit{rectangle with four holes} shape. Similarly, \textit{greedy} and \textit{field-opt-greedy} lay fibers along stress directions, and \textit{field-opt-greedy} provides slightly smoother fiber paths. \textit{Optimized} lays the first fiber over all the holes, and the second fiber around the middle two holes, with paths tightly around the holes.}
    \label{fig:rectangle_4_holes_fiber_paths}
\end{figure*}

%% file: figText/rectangle_4_holes_plot.tex
\begin{figure*}[t]
    \centering
    \begin{subfigure}{0.45\linewidth}
        \includegraphics[width=\linewidth]{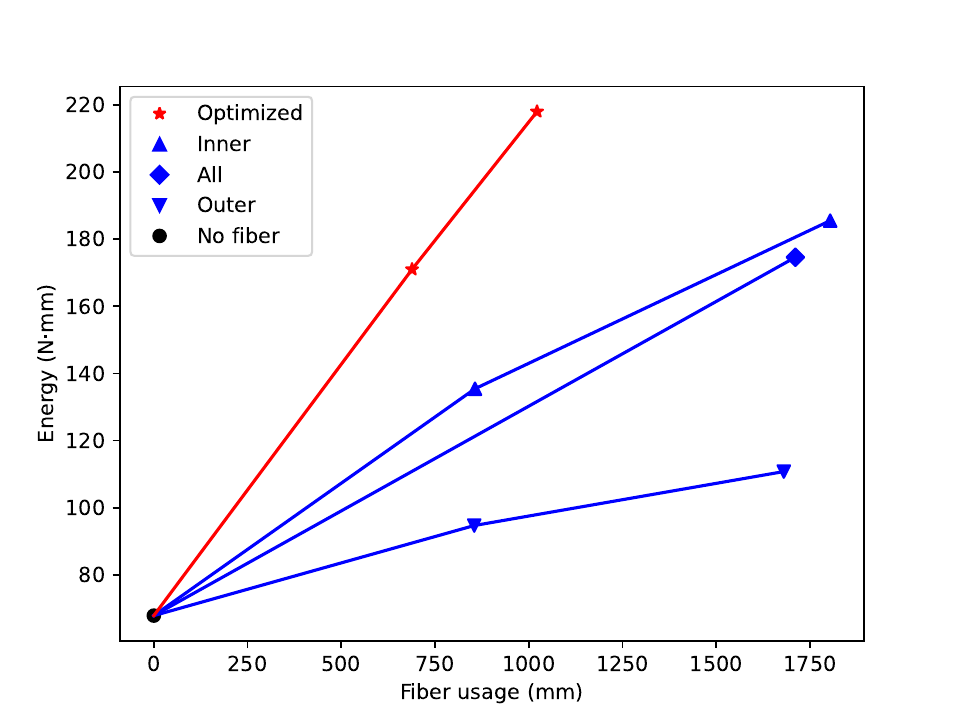}
        \caption{Concentric \textit{vs.} \textit{optimized}}
    \end{subfigure}
    \begin{subfigure}{0.45\linewidth}
        \includegraphics[width=\linewidth]{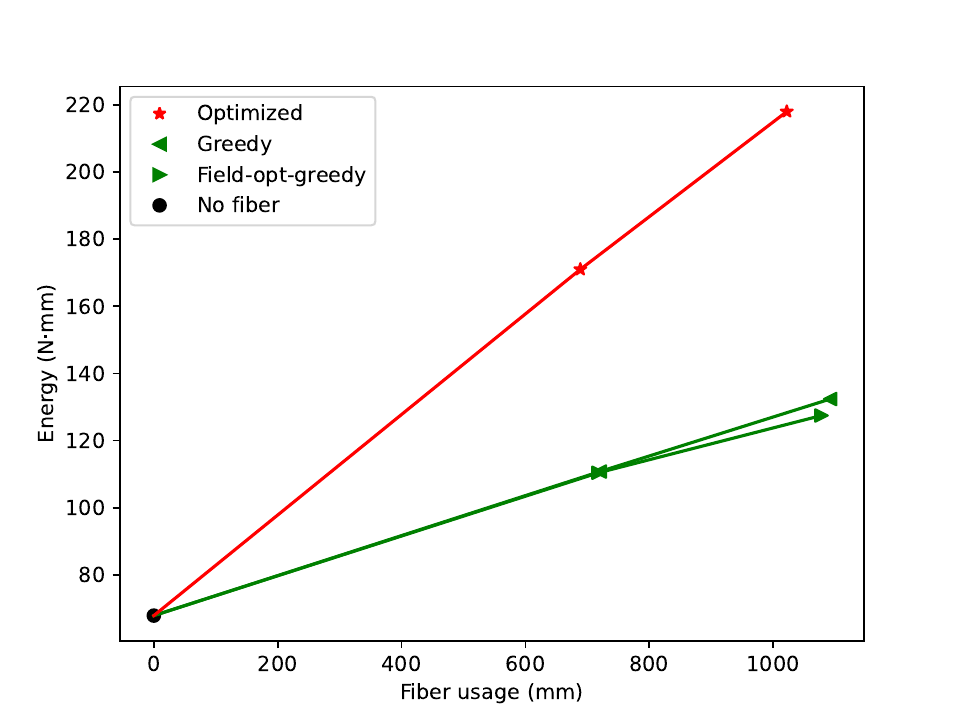}
        \caption{Greedy-based \textit{vs.} \textit{optimized}}
    \end{subfigure}
    \caption{Energy-fiber usage plot (at 1 mm displacement) of all methods on the \textit{rectangle with four holes} shape. The comparison between concentric fiber rings and \textit{optimized} is shown on the left, and the comparison between greedy-based baselines and \textit{optimized} is shown on the right. \textit{Optimized} improves the Pareto front of all the baselines by laying fibers tightly around the holes.}
    \label{fig:rectangle_4_holes_plot}
\end{figure*}

%% file: figText/additional_shapes.tex
\begin{figure*}[t]
    \centering
    \includegraphics[width=\linewidth]{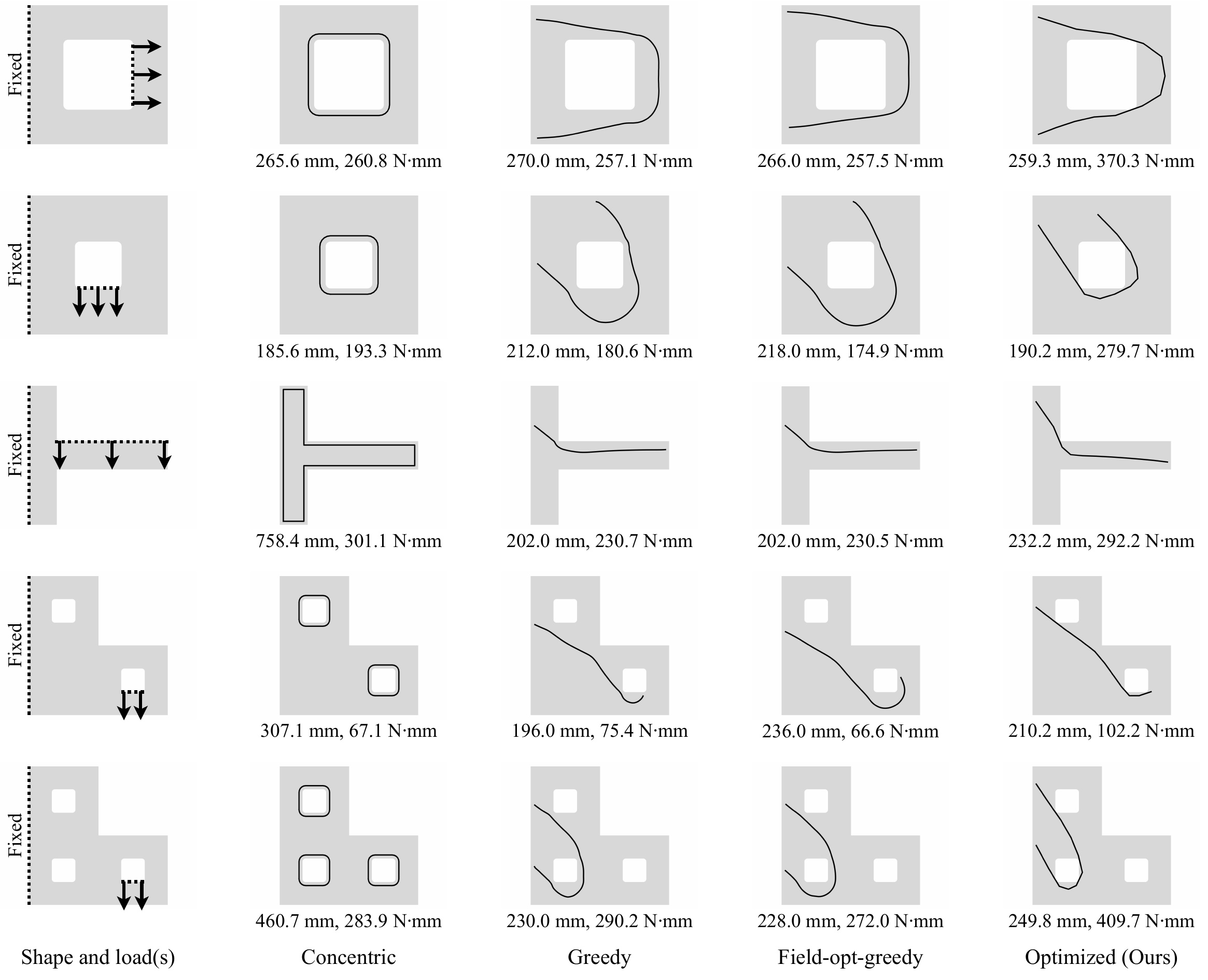}
    \vspace{-10pt}
    \caption{Fiber paths and energy at 1 mm displacement of all methods on additional shapes. Every dotted line indicates a Dirichlet boundary condition. For the first shape (from top to bottom), \textit{optimized} achieves significantly higher energy while using slightly less fiber than all baselines. For the second shape, \textit{optimized} achieves higher energy while using a similar amount of fiber as \textit{concentric} and less fiber than \textit{greedy} and \textit{field-opt-greedy}. For the third shape, \textit{optimized} saves approximately 70\% of fiber usage while achieving similar energy as \textit{concentric}. It also achieves much higher energy than \textit{greedy} and \textit{field-opt-greedy} while using slightly more fiber. For the last two shapes, compared to other baselines, \textit{optimized} achieves significantly higher energy while using a less or comparable amount of fiber.}
    \label{fig:additional_shapes}
\end{figure*}

%% file: text/ablation.tex
\section{Ablation studies}

\label{sec:ablation}

Our algorithm without optimization has been studied in~\sect{sec:experiments} as the \textit{greedy} baseline.
In this section, we study the effects of removing two other components of our method: the Laplacian regularizer and the multi-resolution optimization, using the shape \textit{rectangle with two holes} (\fig{fig:rectangle_2_holes_shape}).

\subsection{Ablation study of the Laplacian regularizer}
\input{figText/ablation_lap}

As both the minimum-length regularizer and the boundary regularizer are intuitively necessary for fiber paths to be long enough for printing purposes and within the object boundary, we study the effect of removing the Laplacian regularizer from the optimization.
We run our algorithm with the same hyper-parameter setting except for $w_\text{lap}=0$. We extract one fiber path and upsample for one time.
As shown in~\fig{fig:ablation_lap}, the optimizer successfully optimizes the low-resolution path as the number of points is still small (\fig{fig:ablation_lap_1}), but introduces jagged results with more degrees of freedom (\fig{fig:ablation_lap_2}), demonstrating the need for some form of regularization.

\subsection{Ablation study of multi-resolution optimization}
\input{figText/ablation_res}

In this subsection, we study how the multi-resolution approach speeds up the optimization process.
For the multi-resolution case, we extract one fiber path, downsample its resolution by a factor of 20, optimize it, and upsample and optimize it three times, with every optimization limited to 100 iterations.
For the single-resolution case, we also extract one fiber path, downsample its resolution by a factor of 2, optimize it and limit the maximum number of optimization iterations to 400.
For a fair comparison, we use the same random seed for both cases when sampling starting points of the greedy path extraction algorithm.
As shown in~\fig{fig:ablation_res}, both cases get similar fiber paths with similar strain energy, but multi-resolution optimization reduces the running time by approximately 40\%.

%% file: figText/ablation_lap.tex
\begin{figure}[t]
    \centering
    \begin{subfigure}{0.49\linewidth}
        \includegraphics[width=\linewidth]{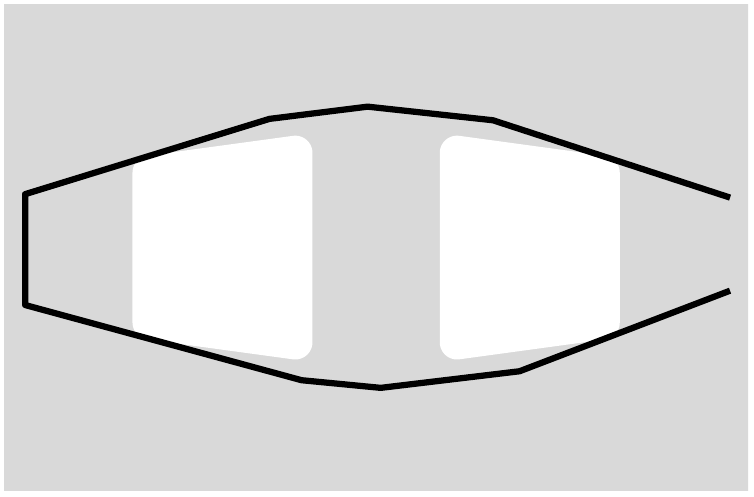}
        \caption{Optimized low-resolution path}
        \label{fig:ablation_lap_1}
    \end{subfigure}
    \begin{subfigure}{0.49\linewidth}
        \includegraphics[width=\linewidth]{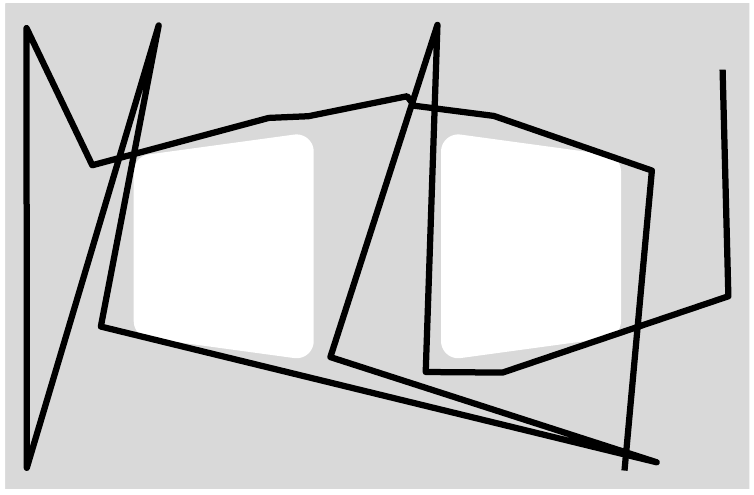}
        \caption{Optimized upsampled path}
        \label{fig:ablation_lap_2}
    \end{subfigure}
    \caption{Optimization results with the Laplacian regularizer disabled. As shown on the left, the optimizer successfully optimizes the low-resolution path. It fails to optimize the fiber path after upsampling, as shown on the right.}
    \label{fig:ablation_lap}
\end{figure}

%% file: figText/ablation_res.tex
\begin{figure}[t]
    \centering
    \begin{subfigure}{0.49\linewidth}
        \includegraphics[width=\linewidth]{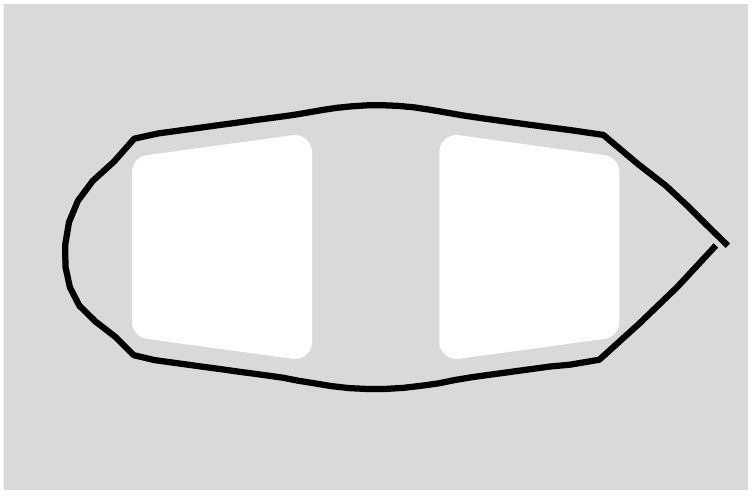}
        \caption{Multi-res, 293 s, 195.8 N$\cdot$mm}
    \end{subfigure}
    \begin{subfigure}{0.49\linewidth}
        \includegraphics[width=\linewidth]{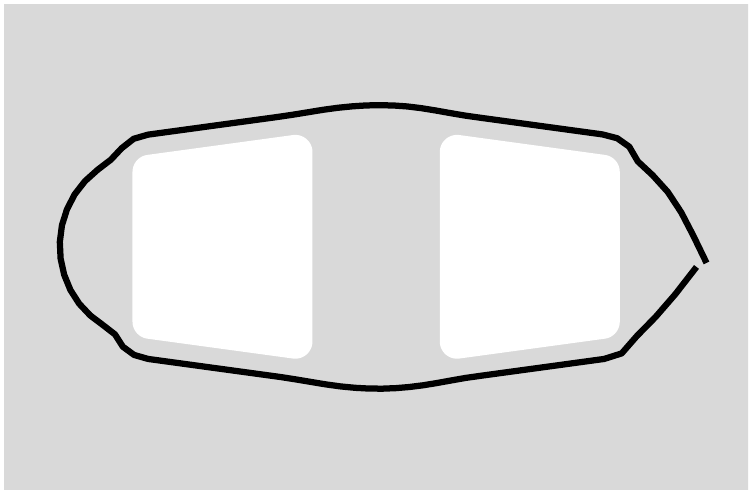}
        \caption{Single-res, 485 s, 190.0 N$\cdot$mm}
    \end{subfigure}
    \caption{Running time and the strain energy at 1 mm displacement for both single-resolution and multi-resolution optimization. In this case, we save approximately 40\% of running time by multi-resolution optimization.}
    \label{fig:ablation_res}
\end{figure}

%% file: text/discussion.tex
\section{Discussion}

In this work, we studied the task of fiber path planning in 3D printing for given external loads, aiming at maximizing the stiffness.
We proposed an end-to-end optimization approach that optimizes regularized object stiffness directly to the fiber layout, rather than an intermediate fiber orientation field, with the help of the adjoint method and automatic differentiation.
We perform planning by extracting fiber paths using a greedy algorithm that lays fiber paths along stress directions, followed by coarse-to-fine optimization.
To apply our method, we first measure the effective moduli of plastic and fiber by manufacturing and testing real 3D prints.
We then study our method with three baselines on four case studies and several additional shapes.
The first baseline is concentric fiber rings from Eiger, a leading digital manufacturing software package developed by Markforged.
The second baseline is our method with the optimization part removed, producing fiber paths from the greedy path extraction algorithm.
The third baseline includes a fiber field optimization part which smooths the stress field before using it in the greedy algorithm.
We demonstrated that, both in simulation and real experiments, our method could generate shorter fiber paths while achieving greater stiffness (\ie, we improved the Pareto front).
We also studied the effects of removing the Laplacian regularizer and the multi-resolution optimization, showing the Laplacian regularizer is necessary for the optimization to be stable and multi-resolution optimization helps reduce the running time.

We would also like to mention some limitations of our method.
First, our simulation simplifies the task by assuming linear elasticity, restricting to in-plane stress, and treating both plastic and fiber as isotropic materials with different Young's moduli and identical Poisson's ratio.
Lifting these assumptions would introduce greater mathematical complexity, but would require no conceptual changes to our approach.
Additionally, the planning is not performed in real time. For example, to plan fiber paths for the shape \textit{rectangle with two holes}, our method uses 10 minutes and 18 minutes to generate the two studied solutions, respectively.
Relative to the time required to design and print a part, this represents only a small increase.
In addition, the hyper-parameters may have to be tuned when the task changes. For example, if we switch to a much larger shape, the scale of strain energy and the lengths of fiber paths will change. We may have to adjust the weight of the Laplacian regularizer, balancing the optimization stability and the variety of fiber paths, though this is usually easy to tune in a few tries.
Lastly, as our optimizer is gradient-based, the optimization may be trapped in a local minimum. Thus a good initialization is important for our method, and we may have to sample greedy paths several times to obtain a good one.

%% file: text/ack.tex
\begin{acks}

We would like to thank Amit Bermano and members of the Princeton Laboratory for Intelligent Probabilistic Systems for valuable discussions and feedback, Joseph Vocaturo and the Princeton Department of Civil and Environmental Engineering for providing the universal testing machine, as well as Markforged.
This work is partially supported by the Princeton School of Engineering and Applied Science, as well as the U.\,S.~National Science Foundation under grants \#IIS-1815070, \#IIS-2007278, and \#OAC-2118201.

\end{acks}

%% file: text/appendix.tex
\appendix
\section{Appendix: field optimization}

Given a stress field $\bm{\sigma}$, we would like to find a fiber field $\bm{v}: \Omega \to \mathbb{R}^2$ such that (1) its direction is aligned with $\bm{\sigma}$; (2) it is smooth. We solve $\bm{v}$ by minimizing an objective function that reflects both proprieties:
\begin{equation}
    \mathcal{L}(\bm{v}; \bm{\sigma}) \coloneqq \alpha_{\texttt{stress}} \cdot \mathcal{L}_{\texttt{stress}}(\hat{\bm{v}}; \bm{\sigma}) + \alpha_{\texttt{smooth}} \cdot \mathcal{L}_{\texttt{smooth}}(\hat{\bm{v}}),
\end{equation}
where $\alpha_{\texttt{stress}}$ and $\alpha_{\texttt{smooth}}$ are hyper-parameters, and 
\begin{equation}
    \hat{\bm{v}}(x, y) \coloneqq \bm{v}(x, y) / ||\bm{v}(x, y)||
\end{equation}
is the normalized $\bm{v}$, as the objective function should be invariant regardless of the length of $\bm{v}(x, y)$. Note that the objective function should also be invariant if we randomly flip some $\bm{v}(x, y)$'s, which needs some special handling, as we will discuss below.

\paragraph{Consistent with $\bm{\sigma}$.}
For a specific point $(x, y) \in \Omega$, we calculate the tension in the stress field $\bm{\sigma}$ along $\hat{\bm{v}}(x, y)$, which is
\begin{equation}
    \hat{\bm{v}}(x, y)^\intercal \bm{\sigma}(x, y) \hat{\bm{v}}(x, y).
\end{equation}
We then integrate it over $\Omega$ and get
\begin{equation}
    \mathcal{L}_{\texttt{stress}}(\hat{\bm{v}}; \bm{\sigma}) \coloneqq -\iint_\Omega \hat{\bm{v}}(x, y)^\intercal \bm{\sigma}(x, y) \hat{\bm{v}}(x, y) \mathrm{d} x \mathrm{d} y,
\end{equation}
where the negative sign indicates we would like to maximize the tension along the field direction.

\paragraph{Smoothness.}
We penalize the squared Frobenius norm of the gradient of $\hat{\bm{v}}$:
\begin{equation}
    \mathcal{L}_{\texttt{smooth}}(\hat{\bm{v}}) \coloneqq \iint_\Omega || \nabla \hat{\bm{v}}(x, y) ||_F^2 \mathrm{d} x \mathrm{d} y.
\end{equation}
Note that the penalty should be invariant to flips of $\hat{\bm{v}}(x, y)$'s, so we handle this invariance when calculating the finite difference:
\begin{align}
\footnotesize
\begin{split}
    || \nabla \hat{\bm{v}}(x, y) ||_F^2 \coloneqq & \min \left( \left| \left| \frac{\hat{\bm{v}}(x + h, y) - \hat{\bm{v}}(x, y)}{h} \right| \right|^2, \left| \left| \frac{\hat{\bm{v}}(x + h, y) + \hat{\bm{v}}(x, y)}{h} \right| \right|^2 \right) \\
    & + \min \left( \left| \left| \frac{\hat{\bm{v}}(x, y + h) - \hat{\bm{v}}(x, y)}{h} \right| \right|^2, \left| \left| \frac{\hat{\bm{v}}(x, y + h) + \hat{\bm{v}}(x, y)}{h} \right| \right|^2 \right),
\end{split}
\end{align}
where $h$ is the step size.

In the experiments, we set $\alpha_{\texttt{stress}}$ to $1$ and $\alpha_{\texttt{smooth}}$ to $0.02$. We use the BFGS optimizer with a gradient tolerance of $1 \times 10^{-6}$ and set the maximum number of iterations to $100$.